\documentclass[aps,prb,longbibliography,twocolumn,superscriptaddress]{revtex4-2}
\usepackage{amssymb,amsmath}
\PassOptionsToPackage{hyphens}{url} % url is loaded by hyperref
\usepackage[unicode=true]{hyperref}
\PassOptionsToPackage{usenames,dvipsnames}{color} % color is loaded by hyperref
\hypersetup{
            pdftitle={Increased localization of Majorana modes in antiferromagnetic chains on superconductors},
            colorlinks=true,
            linkcolor=blue,
            citecolor=blue,
            urlcolor=blue,
            breaklinks=true,
			final=true}
\usepackage{natbib}
\usepackage{graphicx,grffile}
\makeatletter
\def\maxwidth{\ifdim\Gin@nat@width>\linewidth\linewidth\else\Gin@nat@width\fi}
\def\maxheight{\ifdim\Gin@nat@height>\textheight\textheight\else\Gin@nat@height\fi}
\makeatother
\setkeys{Gin}{width=\maxwidth,height=\maxheight,keepaspectratio}
\providecommand{\tightlist}{%
  \setlength{\itemsep}{0pt}\setlength{\parskip}{0pt}}
\newcommand{\sgn}{\text{sign}}
\newcommand{\Pf}{\text{Pf}}

\newcommand{\hc}{\text{H.c.}}
\newcommand{\pd}{{\phantom{\dagger}}}
\renewcommand{\vec}[1]{\mathbf{#1}}

\begin{document}
\title{%{\color{red}\bf v0.7:~}
Increased localization of Majorana modes in antiferromagnetic chains on superconductors }

\author{Daniel Crawford}
\affiliation{School of Physics, University of Melbourne, Parkville, VIC
3010, Australia}
\author{Eric Mascot}
\affiliation{School of Physics, University of Melbourne, Parkville, VIC
3010, Australia}
\author{Makoto Shimizu}
\affiliation{Department of Physics, Okayama University, Okayama
700-8530, Japan}
\author{Roland Wiesendanger}
\affiliation{Department of Physics, University of Hamburg, D-20355 Hamburg, Germany}
\author{Dirk K. Morr}
\affiliation{University of Illinois at Chicago, Chicago, IL 60607, USA}
\author{Harald O. Jeschke}
\affiliation{Research Institute for Interdisciplinary Science, Okayama
University, Okayama 700-8530, Japan}
\author{Stephan Rachel}
\affiliation{School of Physics, University of Melbourne, Parkville, VIC
3010, Australia}

\begin{abstract}
Magnet-superconductor hybrid (MSH) systems are a key platform for
custom-designed topological superconductors. Ideally, the ends of a
one-dimensional MSH structure will host Majorana zero-modes (MZMs), the
fundamental unit of topological quantum computing. However, some of the experiments
with ferromagnetic chains show a more complicated picture. Due to tiny
gap sizes and hence long coherence lengths MZMs might hybridize and
lose their topological protection. Recent experiments on a niobium
surface have shown that both ferromagnetic and antiferromagnetic chains
may be engineered, with the magnetic order depending on the crystallographic
direction of the chain.
While ferromagnetic chains are well
understood, antiferromagnetic chains are less so. Here we study two
models inspired by the niobium surface: a minimal model to elucidate the
general topological properties of antiferromagnetic chains, and an
extended model to more closely simulate a real system by mimicking the proximity effect. 
We find that in general for antiferromagnetic chains the topological gap is larger than for 
ferromagnetic ones and thus coherence lengths are shorter for antiferromagnetic chains, 
yielding more pronounced localization of MZMs in these chains.
While for some parameters antiferromagnetic chains may be topologically trivial,
we find in these cases that adding an additional adjacent chain can result in a nontrivial system, with a single MZM at each chain end.
\end{abstract}

\maketitle

\hypertarget{introduction}{%
\section{Introduction}\label{introduction}}

One-dimensional (1D) topological superconductors (TSCs) are candidates for
hosting Majorana zero-modes (MZMs)\,\citep{kitaev_unpaired_2001, clarke_majorana_2011, deng_anomalous_2012}.
These quasiparticles obey non-Abelian statistics and may be used for
topological, (\emph{i.e.}, fault-tolerant) quantum computing\,\citep{ivanov_non-abelian_2001, nayak_non-abelian_2008}. 1D TSCs can be
engineered --- there are a myriad of proposals\,\citep{lutchyn_majorana_2010, mourik_signatures_2012, pientka_topological_2013, choy_majorana_2011, li_topological_2014, nadj-perge_proposal_2013, klinovaja_topological_2013, martin_majorana_2012, schecter_self-organized_2016}
--- although to date there has not been any completely unambiguous
experimental realization. Magnet-superconductor hybrid (MSH) structures
constitute a particularly promising platform for MZMs, which involve
depositing chains or islands of magnetic adatoms on the surface of a
superconductor by self-assembly or single-atom manipulation using a
scanning-tunneling microscope (STM)\,\citep{choy_majorana_2011, nadj-perge_proposal_2013, li_topological_2014, nadj-perge_observation_2014, schecter_self-organized_2016, kim_toward_2018}.
STM techniques allow for both atomic-scale control of structures and also
atomic-resolution measurements such as spectroscopy\,\citep{ruby_end_2015, palacio-morales_atomic-scale_2019}, reconstruction
of density of states\,\citep{pawlak_probing_2016}, and spin-polarized
maps\,\citep{schneider_atomic-scale_2021}. A rapidly growing number of
MSH systems have been studied in recent years\,\citep{nadj-perge_observation_2014, ruby_end_2015, pawlak_probing_2016, li_two-dimensional_2016, feldman_high-resolution_2017, jeon_distinguishing_2017, ruby_exploring_2017, kim_toward_2018, palacio-morales_atomic-scale_2019, schneider_controlling_2020, odobesko_observation_2020, schneider_atomic-scale_2021, schneider_topological_2021, crawford_majorana_2021, kuster_long_2021, kuster_correlating_2021, brinker_anomalous_2022, kuster_non-majorana_2022, schneider_precursors_2022}.

The first MSH experiment\,\citep{nadj-perge_observation_2014} involved Fe
chains on Pb(110), with the chains grown via self-assembly. The authors
observed signature zero-energy end states, demonstrating the viability
of the platform. Subsequent experiments replicated these results\,\citep{jeon_distinguishing_2017} while
reducing disorder\,\citep{pawlak_probing_2016} and increasing spectral resolution\,\citep{ruby_end_2015, feldman_high-resolution_2017}.
The state of the art progressed by transitioning from self-assembled
chains to artificially constructed Fe chains on Re(0001) using a STM tip\,\citep{kim_toward_2018}; because these chains are constructed
atom-by-atom they are crystalline and disorder-free.
Alongside these developments, there have also been first attempts to engineer 2D structures involving
a Pb/Co/Si(111) heterostructure\,\cite{menard_two-dimensional_2017} and
Fe islands on Re(0001)-O(2$\times$1)\,\citep{palacio-morales_atomic-scale_2019} which showed compelling signatures of chiral Majorana
modes.

Because Nb is the elemental superconductor with highest transition temperature at ambient pressure and has a relatively large
spectral gap of 1.51\,meV, it should be an ideal MSH substrate. Only recently has it
been possible to prepare a sufficiently clean Nb(110) surface. The first
Nb(110) experiments studied single Fe adatoms\,\citep{odobesko_observation_2020}, which was rapidly followed up by Mn
chains\,\citep{schneider_topological_2021}. In the latter experiment
there was sufficient spectral resolution to observe in-gap
Yu-Shiba-Rusinov bands\,\citep{balatsky_impurity-induced_2006} and to
identify a signature topological band inversion. Point-like zero-energy end states
are not observed in these Mn/Nb(110) systems but instead a periodic
accumulation of spectral weight along the sides of the chain, dubbed
\emph{side features}; similar features were also observed in Fe/Nb(110)
systems\,\citep{crawford_majorana_2021}. These are identified as hybridized Majorana modes, and have been proposed to have the same origin\,\citep{crawford_majorana_2021} as the previously observed {\it double eye} feature\,\cite{feldman_high-resolution_2017}. Cr chains on Nb(110) have also been
studied, with no signs of MZMs\,\citep{kuster_long_2021, kuster_correlating_2021, kuster_non-majorana_2022, brinker_anomalous_2022}.

Thus far most theoretical and experimental work has focused on chains
with ferromagnetic (FM) order. Most simulations of 1D MSH systems are
also usually based on simple models which couple magnetic and
superconducting orbitals, and only for one-dimensional structures.
Realistic systems of course consist of many atoms, involving \(s\)-,
\(p\)-, and \(d\)-orbitals, and are constructed from a large 3D
superconducting substrate with a (short) chain deposited somewhere on the
surface. Realistic conventional superconductors also have small spectral
gaps (typically \(<2\) meV), while toy models consider gap sizes of
hundreds of meV. Some experiments\,\citep{nadj-perge_observation_2014}
appear to be consistent with these very simple models featuring point-like MZMs, but later
experiments revealed a more complex spatial structure of the low-energy modes\,\citep{feldman_high-resolution_2017, schneider_topological_2021}. It
seems that more realistic models, possibly based on \emph{ab initio}
methods\,\citep{crawford_majorana_2021}, are necessary to capture all
relevant details of MSH systems.

%%%%%%%%%%%%%%%%%%%%%%%%%%%%%%%%%%%%%%%%%%%%%%%%%%%%%%%%%%%%%%%%%%%%%%%%%%%
\begin{figure}[t]
\centering
\includegraphics{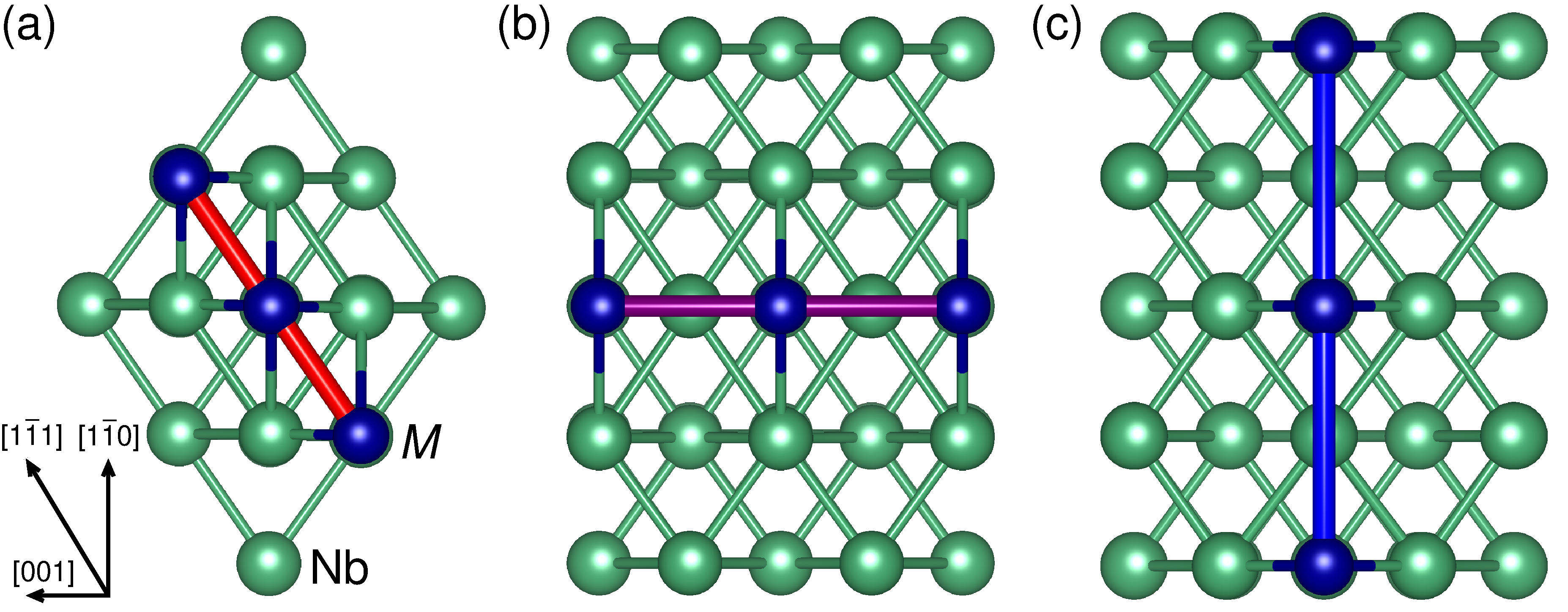}
\caption{Geometry of magnetic chains (\(M\)=Mn, Fe, Cr) on Nb(110)
surfaces. Distances between magnetic ions are determined by the
substrate as (a) \(d_{M-M}^{[1\bar{1}1]}=2.86{\rm \AA}\), (b)
\(d_{M-M}^{[001]}=3.30{\rm \AA}\), and (c)
\(d_{M-M}^{[1\bar{1}0]}=4.51{\rm \AA}\).}
 \label{fig:supercells}
\end{figure}
%%%%%%%%%%%%%%%%%%%%%%%%%%%%%%%%%%%%%%%%%%%%%%%%%%%%%%%%%%%%%%%%%%%%%%%%%%%

Ferromagnetism is not essential for realizing MZMs in MSH chains. Simple single-band models of antiferromagnetic (AFM) chains show MZMs\,\citep{heimes_majorana_2014,heimes_interplay_2015}, as do antiferromagnetic nanowires\,\citep{kobialka_majorana_2021} and superconducting helical magnets\,\citep{martin_majorana_2012}. Not only is the presence of MZMs invariant
to the specific magnetic ground state, but for instance classical Monte-Carlo methods show that FM,
AFM, and spin spiral ground states exist in MSH chains\,\citep{heimes_interplay_2015,neuhaus-steinmetz_complex_2022}. Inspired by these results we
present density functional theory (DFT) calculations for Mn, Fe,
and Cr chains on Nb(110), oriented along three crystalline directions.
We find that exchange couplings between the magnetic moments may be FM
or AFM depending on the type of adatom and on the chain direction. Indeed, recent
experiments\,\citep{beck_spin-orbit_2021, schneider_atomic-scale_2021, lo_conte_coexistence_2022}
have found that Mn atoms deposited on the surface of Nb(110) are
ferromagnetic along the {[}001{]} direction and antiferromagnetic along
the {[}\(1\bar{1}1\){]} direction. In these experiments spin-polarized
STM is used to measure the differential conductance of Mn ultrathin films
and chains. Applying a soft out-of-plane magnetic field to the magnetic
sample reveals FM or AFM order in different directions.

In this work we investigate the spectral and topological properties of
chains of magnetic adatoms on the surface of a conventional
superconductor motivated by the recent experimental developments in Nb-based MSH structures. Depending on the crystalline direction we simulate
either ferromagnetic or antiferromagnetic chains. We typically include
the substrate in the simulations, and also focus on small gaps so as to
account for more realistic scenarios. In Sec.\,\ref{dft-modeling} we present
magnetic couplings of adatom chains on Nb(110) derived from DFT. In
Sec.\,\ref{models-and-method} we introduce two tight-binding models
inspired by the Nb(110) surface. In sections
Sec.\,\ref{bulk-topological-properties} we study their topological phase
diagrams. In sections Sec.\,\ref{chains-on-an-extended-substrate},
Sec.\,\ref{n-leg-ladders}, and
Sec.\,\ref{t-junctions-on-an-extended-substrate} we study different
geometries of these models in real space, including searching for side
features. In Sec.\,\ref{discussion} we discuss our results and summarize
the work in Sec.\,\ref{conclusion}.

%%%%%%%%%%%%%%%%%%%%%%%%%%%%%%%%%%%%%%%%%%%%%%%%%%%%%%%%%%%%%%%%%%%%%%%%%%%
%
%
%%%%%%%%%%%%%%%%%%%%%%%%%%%%%%%%%%%%%%%%%%%%%%%%%%%%%%%%%%%%%%%%%%%%%%%%%%%
\hypertarget{dft-modeling}{%
\section{DFT Modeling}\label{dft-modeling}}

%%%%%%%%%%%%%%%%%%%%%%%%%%%%%%%%%%%%%%%%%%%%%%%%%%%%%%%%%%%%%%%%%%%%%%%%%%%
\begin{table}[b]
	\caption{Magnetic exchange energies ($J_i M^2$) and total magnetic moment per transition metal adatom ($M_{TM}$), calculated within GGA and at least $6\times 6\times 6$ $k$ points (the $k$ mesh was not reduced in the slab direction $k_z$). Positive (negative) magnetic exchange energies indicates AFM (FM) order.
}\label{tab:couplings}
\begin{tabular*}{\columnwidth}{@{\extracolsep{\fill}}c|c|c|c}
transition metal & direction &   $J_i M^2$\,(meV$\mu_{\rm B}^2$)&   $M_{\rm TM}$\,$(\mu_{\rm B})$ \\\hline
Mn & $[1\bar{1}1]$ & 27 & 2.7\\
Mn & $[001]$ & -29 & 2.6 \\
Mn & $[1\bar{1}0]$ & -9 & 2.0 \\\hline\hline
%Mn & & & & & \\
Fe & $[1\bar{1}1]$ & -10 & 2.0 \\
Fe & $[001]$ & -26 & 1.8 \\
Fe & $[1\bar{1}0]$ & 23 & 1.7 \\\hline\hline
Cr & $[1\bar{1}1]$ & 10 & 2.6 \\
Cr & $[001]$ & -41 & 2.7 \\
Cr & $[1\bar{1}0]$ & 10 & 2.4 \\\hline\hline
\end{tabular*}
\end{table}
%%%%%%%%%%%%%%%%%%%%%%%%%%%%%%%%%%%%%%%%%%%%%%%%%%%%%%%%%%%%%%%%%%%%%%%%%%%

We employ DFT calculations based on the full potential local orbital
(FPLO) basis set\,\citep{koepernik_full-potential_1999} and generalized
gradient approximation (GGA) exchange correlation functional\,\citep{perdew_generalized_1996} to investigate the magnetic interactions
of transition metal (TM) chains on the Nb(110) surface. For this
purpose, we construct three different supercells with two symmetry
inequivalent TM sites for the three directions \([1\bar{1}{1}]\),
\([001]\), and \([1\bar{1}0]\) on the Nb(110) surface
(Fig.\,\ref{fig:supercells}). These correspond to nearest, next-nearest,
and third-nearest neighbor distances for the TM adatoms (with exchange couplings $J_1$, $J_2$, $J_3$, respectively); equivalently,
these correspond to TM-TM distances 2.86\,\AA, 3.30\,\AA, and 4.51\,\AA,
respectively. This has been found to be the equilibrium position on the
Nb(110) surface both theoretically and experimentally. We use the
projector augmented wave basis as implemented in VASP\,\citep{kresse_ab_1993, kresse_efficiency_1996} to relax the relevant
supercells for each TM considered. We then extract the Heisenberg
exchange interaction by a simple version of DFT energy mapping which is
very successful for insulating quantum magnets\,\citep{hering_phase_2022}
but has been shown to work for metallic systems as well\,\citep{glasbrenner_effect_2015}. For this purpose, we calculate the
energies of FM and AFM states with high precision. Note that due to the
metallic nature of the TM chain on Nb(110) systems, the energy mapping
is not as precise as for insulating magnets. Metallicity has the
consequence that magnetic moments can differ between FM and AFM spin
configurations as well as for different TM distances. Nevertheless, the
approach can give robust information about the sign of the exchange and
about the relative size of the exchanges in chains running along
different directions on the Nb(110) surface. The results of our
calculations are summarized in Table\,\ref{tab:couplings}. While Mn chains
along \([001]\) and \([1\bar{1}0]\) are FM, along the \([1\bar{1}1]\)
direction they are AFM, in agreement with Ref.\,\citep{schneider_atomic-scale_2021, schneider_topological_2021,  lo_conte_coexistence_2022}. We
predict Fe chains to have AFM order along \([1\bar{1}0]\) and FM along
\([1\bar{1}1]\) and \([001]\), and Cr chains to have AFM order along
\([1\bar{1}0]\) and \([1\bar{1}1]\), and FM along \([001]\). Thus we can
choose FM or AFM coupled chains by choosing an appropriate TM adatom and
crystal direction.

%%%%%%%%%%%%%%%%%%%%%%%%%%%%%%%%%%%%%%%%%%%%%%%%%%%%%%%%%%%%%%%%%%%%%%%%%%%
%
%
%%%%%%%%%%%%%%%%%%%%%%%%%%%%%%%%%%%%%%%%%%%%%%%%%%%%%%%%%%%%%%%%%%%%%%%%%%%
\hypertarget{models-and-method}{%
\section{Models and Method}\label{models-and-method}}

%%%%%%%%%%%%%%%%%%%%%%%%%%%%%%%%%%%%%%%%%%%%%%%%%%%%%%%%%%%%%%%%%%%%%%%%%%%
\begin{figure}[t]
\centering
\includegraphics{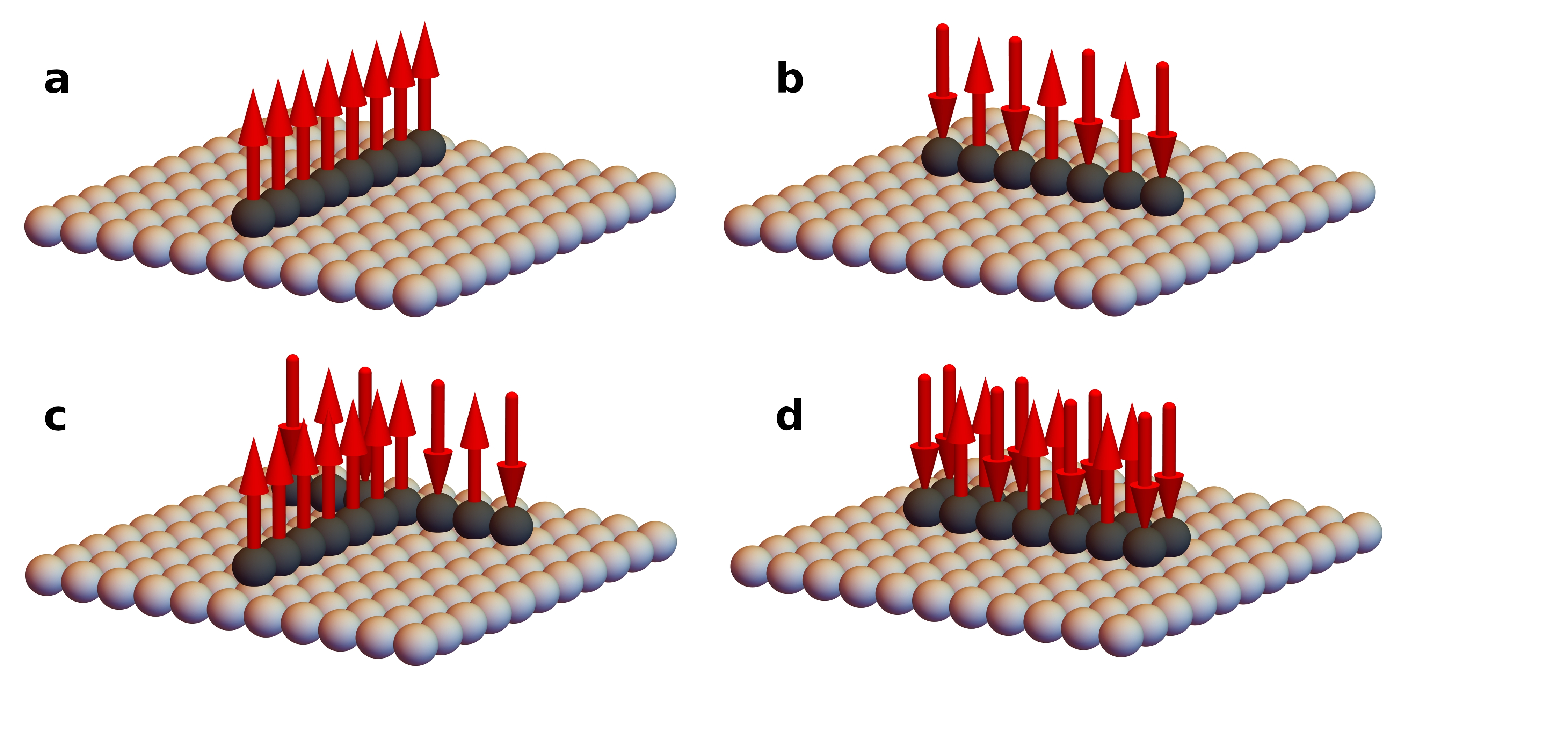}
\caption{Schematics of magnetic adatoms on an Nb-inspired
superconducting substrate. We investigate the following setups: (a) Shiba chain
with FM order; (b) Shiba chain with AFM order; (c) T-junction comprised
of an AFM chain and a FM chain; and (d) AFM two-leg ladder.
\label{fig:setup}}
\end{figure}
%%%%%%%%%%%%%%%%%%%%%%%%%%%%%%%%%%%%%%%%%%%%%%%%%%%%%%%%%%%%%%%%%%%%%%%%%%%

We work with two models inspired by the Nb(110) surface:

\begin{itemize}
\tightlist
\item
  the \emph{minimal model}, in which the unit cell has only one orbital
  on a square lattice. Both superconductivity and magnetic couplings are
  associated with this orbital. In one direction the magnetic couplings
  are AFM and in the other they are FM. Magnetism is restricted to
  chains on a subset of the lattice.
\item
  the \emph{extended model}, in which the unit cell has separate
  superconducting and magnetic sites. Because the substrate is included
  in the unit cell we work with this model only in a 1D geometry. By comparing
  this model to the minimal model we can test for trends in the topological
  physics of AFM chains.
\end{itemize}

For both models we assume a constant order parameter for simplicity; while we can treat the order parameter self-consistently this does not change any of the important physics \citep{awoga_disorder_2017}. We also only consider disorder-free surfaces because FM chains are robust against disorder \citep{awoga_disorder_2017}.

As seen in Fig.\,\ref{fig:supercells}, the \([001]\) and \([1\bar{1}1]\)
directions form a 60° angle with respect to each other. In the minimal model we place the
AFM and FM chains at right angles to each other; this model is concerned
only with the essential physics and so the choice of the relative angles has been neglected in the following.  The model could be extended to the full bcc(110) surface, but this is left for future work.

The 1D FM Bogoliubov--de Gennes (BdG) Hamiltonians we consider possess only particle-hole symmetry so they belong to the D symmetry class in the periodic table of topological classes\,\citep{altland_nonstandard_1997, schnyder_classification_2008, kitaev_periodic_2009}. Thus the relevant topological invariant is a \(\mathbb{Z}_2\) index, Kitaev's Majorana number\,\citep{kitaev_unpaired_2001},

\begin{equation}
\mathcal{M} = \sgn(\Pf[i\tilde{H}(0)]\Pf[i\tilde{H}(\pi)]),
\end{equation}

where \(\tilde{H}(k)\) is the Hamiltonian in the Majorana basis at momentum \(k\).
In contrast, the 1D AFM BdG Hamiltonians we consider possess
particle-hole, chiral, and time-reversal symmetries so they belong to
the BDI symmetry class. In this case the relevant topological invariant is a \(\mathbb{Z}\) index,
the winding number\,\citep{steffensen_topological_2022},

\begin{equation}
\mathcal{W} = \frac{1}{2\pi i} \int_{-\pi}^\pi dk \frac{\partial_k \det[V_k]}{\det[V_k]},
\end{equation}

where \(V_k\) is the off-diagonal block of the Hamiltonian in the
Majorana basis at momentum \(k\). Only when \(\mathcal{W}\) is odd do we
get unpaired MZMs at chain ends, because pairs of MZMs annihilate; in fact, we can relate the two invariants by \(\mathcal{M} = \mathcal{W} \mod 2\)\,\citep{kobialka_majorana_2021}.
Nontrivial phases \(\mathcal{W}\neq 0, \mathcal{M}=-1\) are only valid when there is a bulk spectral gap, and
phase transitions, \emph{i.e.}, changes of $\mathcal{W}, \mathcal{M}$ are associated with a gap closing. Thus by computing
the spectral gap in periodic boundary conditions (PBC) we can confirm a
topological phase via gap closings, and in open boundary conditions
(OBC) we can identify potential topological phases by identifying
extended regions with zero-energy states in parameter space (\emph{i.e.}, via the
bulk-boundary correspondence). When we include the substrate we cannot compute the invariant because the system becomes inhomogenous. In these cases we rely on the bulk-boundary correspondence to
identify topological phases.

%%%%%%%%%%%%%%%%%%%%%%%%%%%%%%%%%%%%%%%%%%%%%%%%%%%%%%%%%%%%%%%%%%%%%%%%%%%
%
%
%%%%%%%%%%%%%%%%%%%%%%%%%%%%%%%%%%%%%%%%%%%%%%%%%%%%%%%%%%%%%%%%%%%%%%%%%%%
\hypertarget{minimal-model}{%
\subsection{Minimal model}\label{minimal-model}}

We start with the prototypical FM Shiba lattice model\,\citep{li_two-dimensional_2016,rachel_quantized_2017,crawford_high-temperature_2020}. There is a single orbital per unit cell
(with spin degree of freedom) which captures both superconductivity and
magnetism. The superconductor is modeled as a two-dimensional square
lattice \(\Lambda\) spanned by \(\hat{\vec{e}}_1\) and
\(\hat{\vec{e}}_2\) with magnetic adatoms occupying a subset
\(\Lambda^* \subseteq \Lambda\). We extend this model to support AFM
order by doubling the unit cell such that there is AFM order in the
\(x\) direction and FM order in the \(y\) (\emph{i.e.}, row-wise AFM
order on a square lattice). We emphasize that though the magnetic unit
cell is doubled, all other parameters remain identical on both sublattices.
For simplicity we work in terms of unit cells so \(N_x=16\) indicates 32
atoms in the \(x\) direction, while \(N_y=32\) indicates 32 atoms in the
\(y\) direction.

The tight-binding BdG Hamiltonian is defined as

\begin{align}
\nonumber
H &= \sum_{\vec{r} \in \Lambda} \big[ t(a_\vec{r}^\dagger b^\pd_\vec{r} + b_\vec{r}^\dagger a^\pd_{\vec{r}+\hat{\vec{e}}_1} + a^\dagger_\vec{r} a^\pd_{\vec{r} + \hat{\vec{e}}_2}  + b^\dagger_\vec{r} b^\pd_{\vec{r} + \hat{\vec{e}}_2}) \\[5pt]
\nonumber
&+ \frac{\mu}{2}(a_\vec{r}^\dagger a^\pd_\vec{r} + b_\vec{r}^\dagger b^\pd_\vec{r}) \\[5pt]
\nonumber
&+ [i \alpha(a_\vec{r}^\dagger \sigma_y b^\pd_\vec{r} + b_\vec{r}^\dagger \sigma_y a^\pd_{\vec{r}+\hat{\vec{e}}_1} - a_\vec{r}^\dagger \sigma_x a^\pd_{\vec{r} + \hat{\vec{e}}_2} - b_\vec{r}^\dagger \sigma_x b^\pd_{\vec{r} + \hat{\vec{e}}_2})] \\[5pt]
\nonumber
&+ \Delta(a_{x,\uparrow}^\dagger a_{x,\downarrow}^\dagger + b_{x,\uparrow}^\dagger b_{x,\downarrow}^\dagger) + \hc \big] \\[5pt]
&+ J \sum_{\vec{r} \in \Lambda^*} (a_\vec{r}^\dagger \sigma_z a^\pd_\vec{r} - b_\vec{r}^\dagger \sigma_z b^\pd_\vec{r}) .
\label{eq:hamiltonian}
\end{align}

Here
\(a_\vec{r}^\dagger = \begin{pmatrix} a_{\vec{r},\uparrow}^\dagger & a_{\vec{r},\downarrow}^\dagger \end{pmatrix}\)
is a spinor of the creation operators for electrons at site \(\vec{r}\)
for sublattice \(a\) with spin \(\uparrow,\downarrow\) (and similarly
for sublattice \(b\)). \(\sigma_{x,y,z}\) are the three Pauli matrices.
\(t\) is the nearest-neighbor hopping amplitude; \(\mu\) the chemical
potential; \(\alpha\) the Rashba spin-orbit coupling strength; and \(J\)
the Zeeman strength resulting from the magnetic moments of the adatoms.
Superconductivity is induced by the proximity effect and has magnitude
\(\Delta\). In Sec.\,\ref{bulk-topological-properties} we study this model
in a 1D AFM (FM) variant by dropping all terms with hopping in \(y\)
(\(x\)) direction.

%%%%%%%%%%%%%%%%%%%%%%%%%%%%%%%%%%%%%%%%%%%%%%%%%%%%%%%%%%%%%%%%%%%%%%%%%%%
%
%
%%%%%%%%%%%%%%%%%%%%%%%%%%%%%%%%%%%%%%%%%%%%%%%%%%%%%%%%%%%%%%%%%%%%%%%%%%%
\hypertarget{extended-models}{%
\subsection{Extended models}\label{extended-models}}

The extended model was introduced as the four-site model in Ref.\,\citep{crawford_majorana_2021}. Originally introduced to mimic the
Mn/Nb(110) surface along the {[}001{]} direction, (\emph{i.e.}, with FM
couplings), here we study a variant with doubled unit cell and AFM Zeeman
couplings. Hence the FM extended model contains three superconducting
atoms and one magnetic adatom per unit cell, while the AFM extended
model contains six superconducting atoms and two magnetic adatoms per
unit cell. The tight-binding Hamiltonian of the normal state is
\begin{align}
\nonumber
H_t = &\sum_{ij a b} t_{ij}^{a b} c_i^{\dagger a} c_j^b + i\alpha_{ij}^{a b} c_i^{\dagger a}\sigma_2c_j^b \\[5pt]
&- \mu\sum_{i a} c_i^{\dagger a} c_i^ a + J\sum_{i, a=0,4} \delta^a c_i^{\dagger  a}\sigma_3 c_i^ a. \label{eq:ham2}
\end{align}
Here \(i,j\) enumerates unit cells; \(a,b\) labels the eight atoms, with \(a=0,4\) indicating the magnetic adatoms;
\(c_i^{\dagger a} = (c_{i,\uparrow}^{\dagger a}, c_{i,\downarrow}^{\dagger a})\) is a spinor of the creation operators at site \(i\) for atom \(a\); and
\(t_{ij}^{ab}\) are the hopping amplitudes between atoms \(a,b\) at sites \(i,j\).
\(\alpha_{ij}^{ab}\) are the Rashba spin-orbit coupling amplitudes between atoms \(a,b\) and sites \(i,j\).
The elements of \(t_{ij}^{ab}\) are given in Appendix\,\ref{extended-model-definition}; \(\alpha_{ij}^{ab}\) is the same as
\(t_{ij}^{ab}\), except with the elements replaced by a single value \(\alpha\).
\(J\) is the magnetic exchange coupling of atoms $a=0,4$, with \(\delta^a = +1\) for \(a=0\) and \(\delta^a = -1\) for \(a=4\). We
add onsite \(s\)-wave superconductivity on the substrate sites:
\begin{equation}
H_\Delta = \Delta \sum_{i a}^{ a\neq 0, 4} c_{i,\uparrow}^{\dagger a} c_{i,\downarrow}^{\dagger a} + \hc.
\end{equation}
Hence the total Hamiltonian is \(H = H_t + H_\Delta\). After performing a Bogoliubov
transformation, we calculate the relevant topological and spectral
properties. We emphasize that the magnetic adatoms ($a=0,4$) are not directly coupled to the superconducting pairing term. Since all the other sites within the unit cell are superconducting, we simulate here the proximity effect: if $\Delta$ is sufficiently large, also the 0 and 4 sites become effectively superconducting due to their {\it proximity} to the other superconducting atoms, to which they are coupled through hopping amplitudes $t_{ij}^{ab}$.

%%%%%%%%%%%%%%%%%%%%%%%%%%%%%%%%%%%%%%%%%%%%%%%%%%%%%%%%%%%%%%%%%%%%%%%%%%%
\begin{figure}[t!]
\centering
\includegraphics[width=1 \columnwidth]{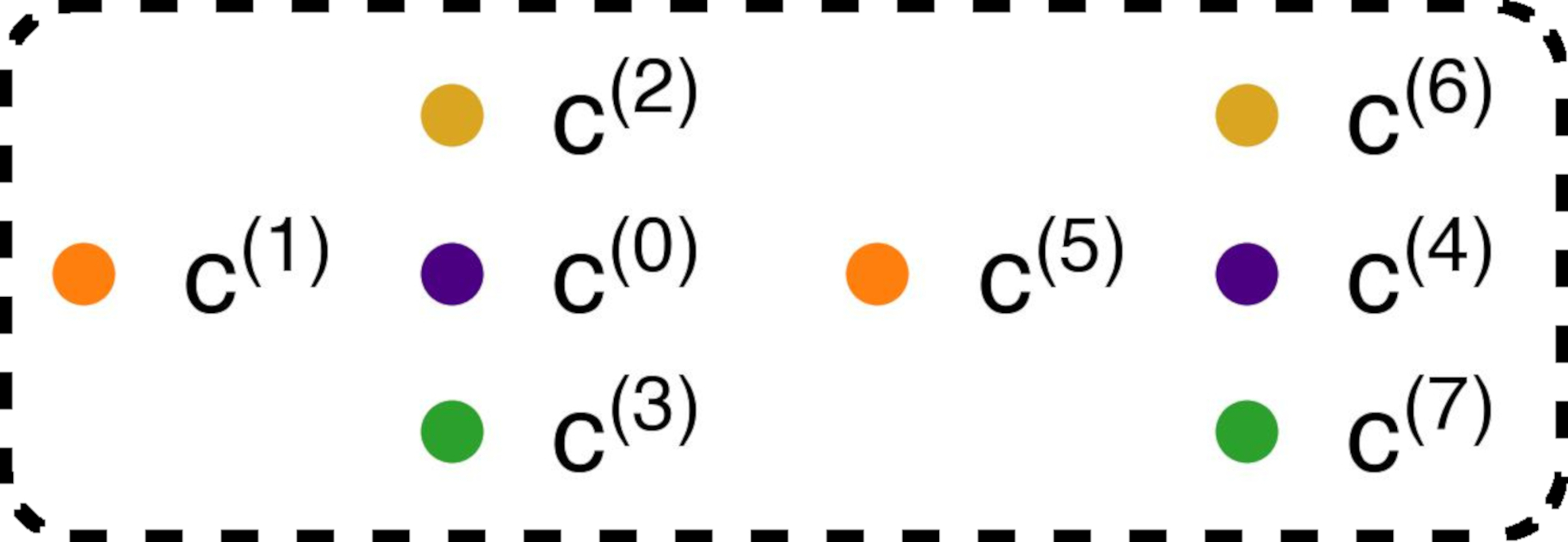}
	\caption{Unit cell for the extended model. \(c^{(0,4)}\) are magnetic adatoms and the others superconducting surface atoms. Note that the unit cell is doubled compared to the ferromagnetic four-site model\,\citep{crawford_majorana_2021}.
\label{fig:eight_site_model}}
\end{figure}
%%%%%%%%%%%%%%%%%%%%%%%%%%%%%%%%%%%%%%%%%%%%%%%%%%%%%%%%%%%%%%%%%%%%%%%%%%%

%%%%%%%%%%%%%%%%%%%%%%%%%%%%%%%%%%%%%%%%%%%%%%%%%%%%%%%%%%%%%%%%%%%%%%%%%%%
%
%
%%%%%%%%%%%%%%%%%%%%%%%%%%%%%%%%%%%%%%%%%%%%%%%%%%%%%%%%%%%%%%%%%%%%%%%%%%%
\hypertarget{bulk-topological-properties}{%
\section{Bulk topological
properties}\label{bulk-topological-properties}}

\begin{figure}
\centering
\includegraphics{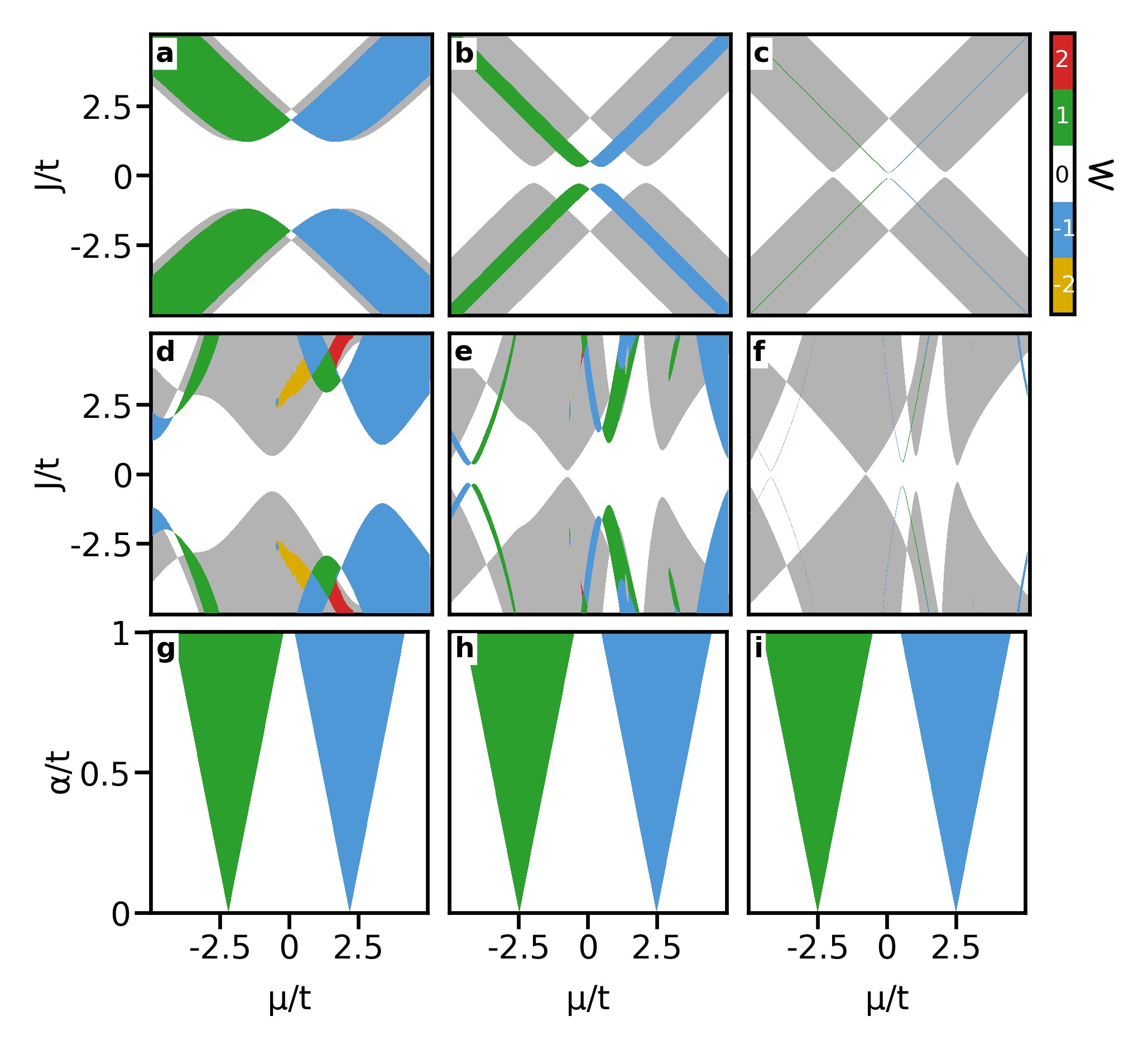}
\caption{Topological phase diagrams as a function of $J$ and $\mu$ for (a-c) minimal
AFM model and (d-f) AFM extended model. Colored regions indicate nontrivial (\(\mathcal{W}\neq  0\)) phases for the AFM variants and gray regions indicate nontrivial (\(\mathcal{M}=-1\)) phases for the FM variants. (g-i) Topological phase diagrams of the minimal AFM model in the purely 1D variant as a function of $\alpha$ and $\mu$.
 Parameters used in (a,d): (\(\Delta\),
\(\alpha\)) = \((1.2, 0.8)t\); in (b,e): (\(\Delta\), \(\alpha\)) =
\((0.3, 0.2)t\); in (c,f): (\(\Delta\), \(\alpha\)) = \((0.1, 0.01)t\);
in (g): (\(\Delta\), \(J\)) = \((1.2, 2.5)t\); in (h): (\(\Delta\), \(J\)) =
\((0.3, 2.5)t\); in (i): (\(\Delta\), \(J\)) = \((0.1, 2.5)t\).
\label{fig:phase_diagrams_collated}}
\end{figure}

We start by studying the topological phase of 1D AFM chains in dependence of $J$ and $\mu$, and compare
to 1D FM chains. Complementary to prior work we find that, for both
minimal and extended models, AFM chains exhibit topologically nontrivial
phases (Fig.\,\ref{fig:phase_diagrams_collated}).
In panels (a-f) we see
that for AFM chains the nontrivial phase shrinks --- that is, takes up a
smaller proportion of parameter space --- as $\Delta$ and $\alpha$ shrink.
For FM chains this trend does not exist (the FM phase diagram is shown in gray in the same panels for comparison). In panels (g-i) we show
topological phase diagrams for the minimal 1D AFM model as a function of $\alpha$ and $\mu$. In agreement with Ref.\,\citep{heimes_interplay_2015}, we find a strong dependence on
\(\alpha\), and only a very weak dependence on \(\Delta\).
For the minimal model we can find the phase boundaries analytically
(see appendix for details; the extended model is not
amenable to the same analytic treatment). These are the hyperbola
\(J^2 = (2\alpha \pm \mu)^2 + \Delta^2\). In contrast, for the 1D FM
variant the phase boundaries are \(J^2 = (\pm 2t -\mu)^2 + \Delta^2\). Equivalently, one can understand the dependence on $\alpha$ as a consequence of the reduced Brillouin zone \citep{heimes_interplay_2015}. For FM chains the phase transition is due to gap closings at $k=0,\pi$ points, whereas for AFM chains the gap closings are at $k=0$ in the reduced Brillouin zone. This corresponds to $k=\pi/2a$ in the original Brillouin zone, with $a$ the lattice spacing.
To summarize, while for both FM and AFM models Rashba spin-orbit coupling (SOC)
is required to realize a topological phase, for AFM models there is a critical
dependence on the SOC magnitude \(\alpha\).

%%%%%%%%%%%%%%%%%%%%%%%%%%%%%%%%%%%%%%%%%%%%%%%%%%%%%%%%%%%%%%%%%%%%%%%%%%%
%%%%%%%%%%%%%%%%%%%%%%%%%%%%%%%%%%%%%%%%%%%%%%%%%%%%%%%%%%%%%%%%%%%%%%%%%%%

\begin{figure}
\centering
\includegraphics{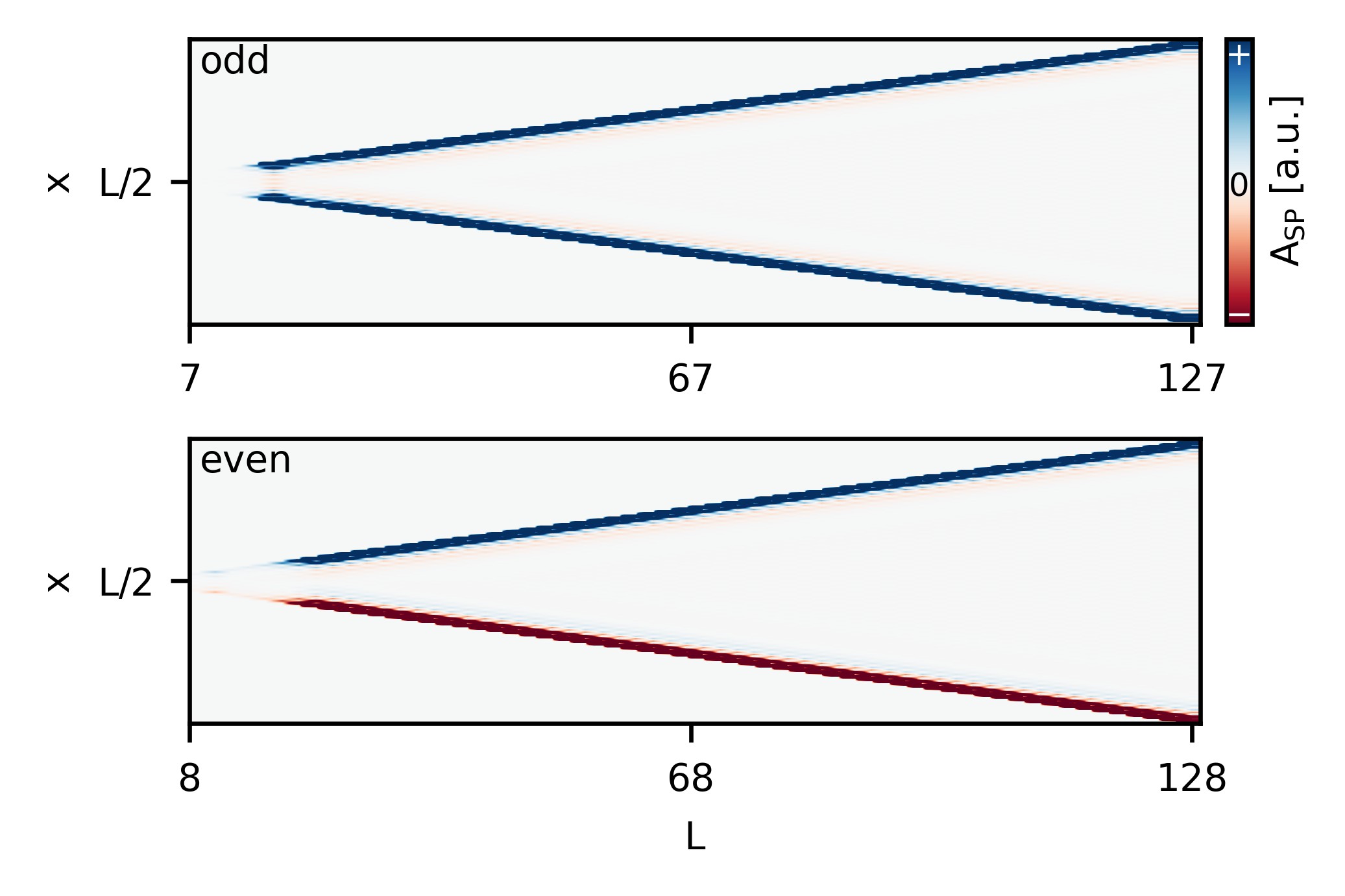}
	\caption{Spin-polarization asymmetry \(A_{\rm SP}\) in the minimal model for odd (top row) and even chain lengths \(L\) (bottom row), with no substrate, in the topological phase. Chains are centered around \(x=64\). For odd chain lengths there is greater spectral weight for the spin-$\downarrow$ component and the spectral weight is symmetric around the center of the chain, similar to FM chains\,\citep{mashkoori_identification_2020, schneider_atomic-scale_2021}. For even chain lengths the chain ends are spin-polarized.
Parameters: (\(\Delta, \alpha, \mu, J) = (1.2, 0.8, 4.5, 4.5)t\).
\label{fig:spin_polarisation}}
\end{figure}

In FM chains the MZMs are in general spin-polarized\,\citep{mashkoori_identification_2020, schneider_atomic-scale_2021}.
We check for a similar effect here by comparing the asymmetry \(A_{\rm SP}\) in the spin-resolved \(E=0\) local density of states (LDOS) for odd and even chain lengths (Fig.\,\ref{fig:spin_polarisation}), with \(A_{\rm SP} = \rho_\downarrow (x) - \rho_\uparrow (x)\) and \(\rho_\sigma (x) \) the \(E=0\) LDOS resolved for spin \(\sigma\) and position \(x\). Fig.\,\ref{fig:spin_polarisation} presents increasing chain lengths from $L=8$ to $L=127$ (top row, odd chains only) and $L=128$ (bottom row, even chains only), with all chains centered around $x=64$; the increased LDOS stemming from the MZMs corresponds to the ends of the chains. For odd chain lengths both ends of the chain are spin-$\downarrow$ terminated. The top row shows greater spectral weight for the spin-$\downarrow$ component of the LDOS compared to the spin-$\uparrow$ \emph{i.e.}, the polarization is aligned with the terminating sites. Because the local Zeeman term points in the same direction as the terminating sites, the spectral weight is symmetric around the center of the chain. In contrast, for even chain lengths the chain ends become spin-polarized, with one end being dominantly spin-$\uparrow$ and the other being spin-$\downarrow$, in accordance with the termination of the local Zeeman field.

In the next section we investigate how the topological phase changes when
the chain is embedded in an extended substrate, and analyze whether the trend of increased localization
of MZMs persists. In Sec.\,\ref{n-leg-ladders} we show a simple geometry
which greatly expands the proportion of parameter space which is topologically
nontrivial.

%%%%%%%%%%%%%%%%%%%%%%%%%%%%%%%%%%%%%%%%%%%%%%%%%%%%%%%%%%%%%%%%%%%%%%%%%%%
%%%%%%%%%%%%%%%%%%%%%%%%%%%%%%%%%%%%%%%%%%%%%%%%%%%%%%%%%%%%%%%%%%%%%%%%%%%

%%%%%%%%%%%%%%%%%%%%%%%%%%%%%%%%%%%%%%%%%%%%%%%%%%%%%%%%%%%%%%%%%%%%%%%%%%%
%
%
%%%%%%%%%%%%%%%%%%%%%%%%%%%%%%%%%%%%%%%%%%%%%%%%%%%%%%%%%%%%%%%%%%%%%%%%%%%
\hypertarget{chains-on-an-extended-substrate}{%
\section{Chains on an extended
substrate}\label{chains-on-an-extended-substrate}}

\begin{figure}
\centering
\includegraphics{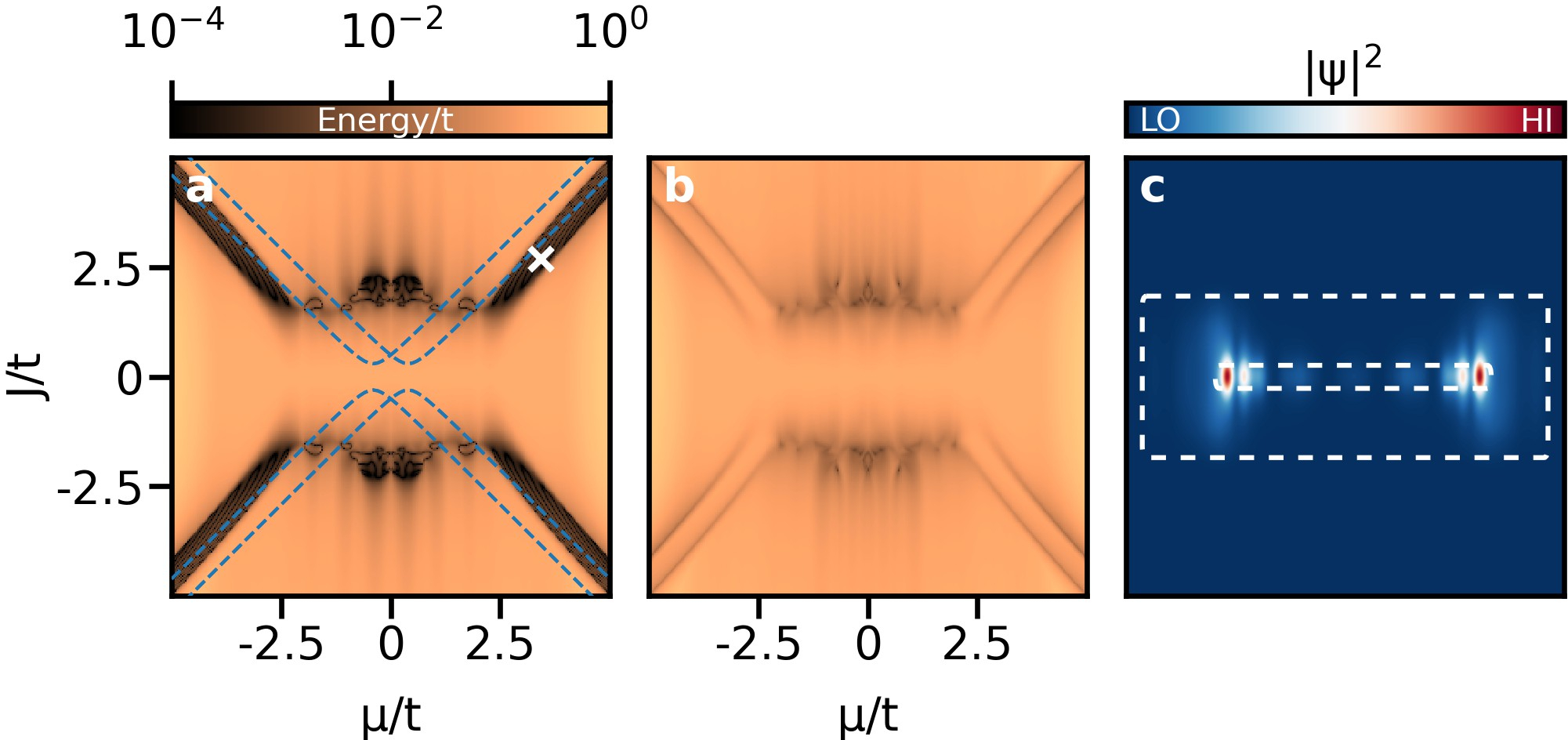}
\caption{First (a) and second (b) positive eigenenergies for the minimal AFM model.
Blue dashed lines in (a) indicate the phase boundaries under the 1D AFM variant
(Fig.\,\ref{fig:phase_diagrams_collated}).
(c) Zero-energy
state corresponding to the parameters as indicated by the white \(\times\) in the gap diagram (a). White dashed box in (c)
indicates the substrate boundaries, and the gray dashed lines the chain
extent. Parameters used for (c): $(J,\mu)=(2.7, 3.3)t$.
Parameters used for all panels: (\(\Delta\), \(\alpha\), \(N_x\), \(N_y\), \(L\)) = \((0.3t, 0.2t, 25, 7, 16)\).
\label{fig:minimal_chain_on_substrate}}
\end{figure}

We now consider AFM chains on an extended, 2D substrate. This is accomplished by including
all \(x,y\) terms in Eq.\,\eqref{eq:hamiltonian} and setting \(\Lambda^*\) to be a 1D region of some
length \(L < N_x\) in the middle of the lattice. While extending the
substrate in this fashion cannot entirely destroy the nontrivial phase,
there may be some changes to the phase diagram. Unfortunately, the
winding number is not defined when the substrate is extended in this
fashion and so we rely on the presence of zero-energy end-states,
protected by a gap, to indicate a nontrivial phase.
Fig.\,\ref{fig:minimal_chain_on_substrate} shows two energy diagrams and a
representative zero-energy state for the minimal model with a small gap.
Panel (a) corresponds to the first positive eigenenergy and panel (b)
corresponds to the second positive eigenenergy. In panel (a) we see dark
regions corresponding to zero-energy states with a similar shape and
extent as that of the 1D AFM phase diagram. In panel (b) we see these
zero-energy states are protected by a gap. Indeed, there are even signs
of gap closing at borders corresponding to the dark regions in (a),
\emph{i.e.}, a phase boundary. Clearly the most stable regions are the diagonal stripes where the energy gap between the zero-energy states and the next-higher energy state is largest.
In panel (c) we show a representative example where the spectral
weight of a zero-energy states is mostly concentrated at chain ends.
Hence we identify these regions as a nontrivial topological phase and
these states as MZMs. The effect of the substrate then is to shift and
distort the topological phase in parameter space. Notably, around \(\mu=0\) the phase boundaries are unclear and the lowest energy state is not well gapped so we identify this region as topologically trivial.

%%%%%%%%%%%%%%%%%%%%%%%%%%%%%%%%%%%%%%%%%%%%%%%%%%%%%%%%%%%%%%%%%%%%%%%%%%%
%%%%%%%%%%%%%%%%%%%%%%%%%%%%%%%%%%%%%%%%%%%%%%%%%%%%%%%%%%%%%%%%%%%%%%%%%%%

Previously we have found that MZMs along {[}001{]} in Mn/Nb(110) systems
appear on the sides of the chain rather than at chain ends, in
contrast to predictions from simplified models\,\citep{crawford_majorana_2021}.
We argue that this occurs due to three complementary factors: (1) A non-trivial hopping structure as realized in multi-orbital systems, (2)
hybridization of the MZMs, and (3) interplay of ferromagnetism and
superconductivity. Hybridization occurs when MZMs overlap, which is
possible when chains are short and coherence lengths are long (for small
gap sizes \(\Delta_{\rm gap} \ll t\) the coherence length can be hundreds of
lattice spacings). This hybridization reveals itself by the absence of end states and instead an oscillating wavefunction\,\cite{crawford_majorana_2021}. Because ferromagnetism is antagonistic to
superconductivity, under certain conditions MZMs can be pushed off the
chain onto the substrate beside it. Taken together, these two factors produce side features. We see this in experiments\,\citep{crawford_majorana_2021, schneider_topological_2021} and
generically in our FM models. In the following, we test whether our AFM models might reveal similar features.

%%%%%%%%%%%%%%%%%%%%%%%%%%%%%%%%%%%%%%%%%%%%%%%%%%%%%%%%%%%%%%%%%%%%%%%%%%%
\begin{figure}[t]
\centering
\includegraphics{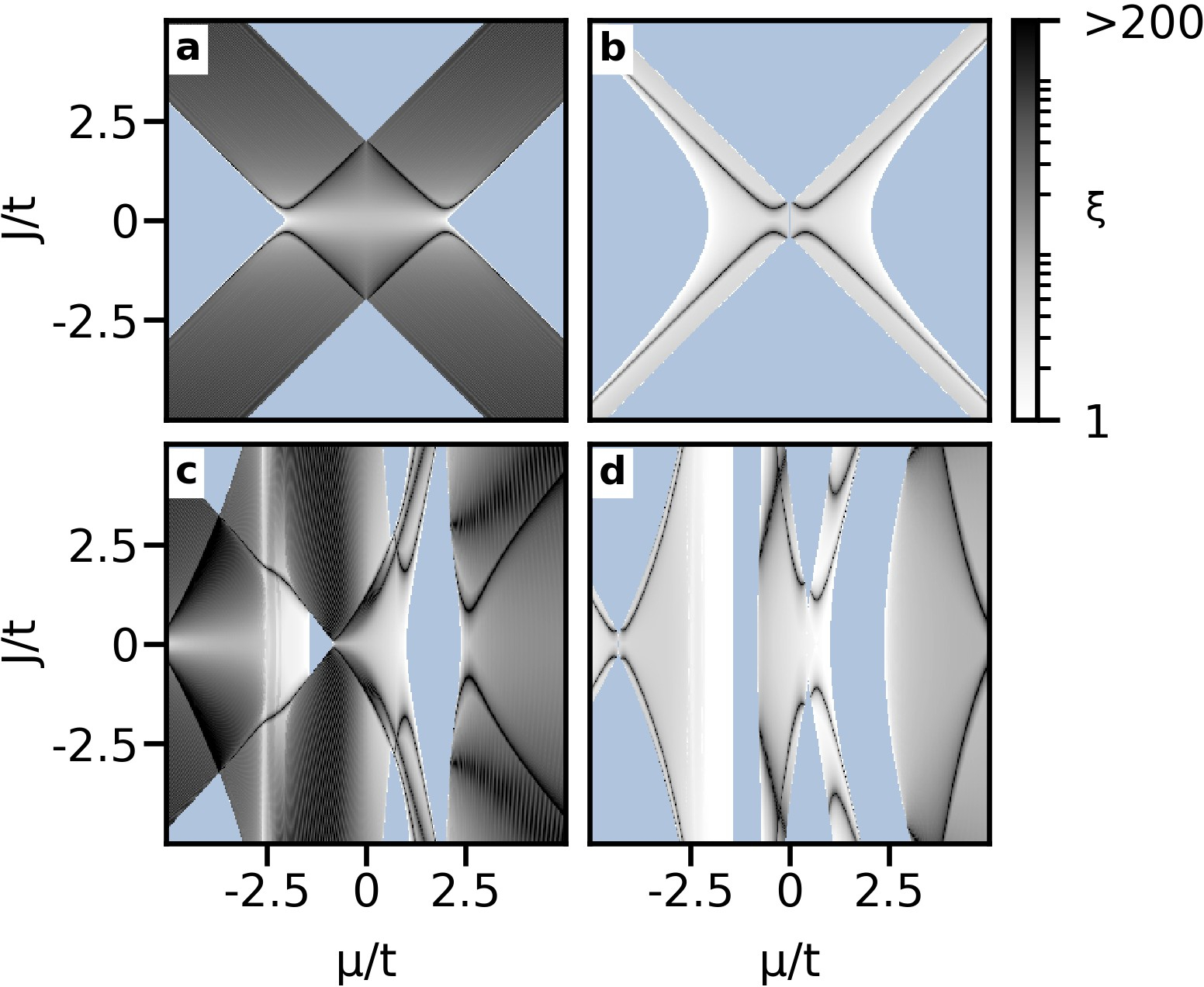}
\caption{Coherence length \(\xi = v_F/\Delta_{\rm gap}\) in the minimal
model for the FM (a) and the AFM (b) variants, and in the extended
model for the FM (c) and AFM (d) variants. Blue regions have undefined
coherence length, \emph{i.e.}, a gapped normal state. In general, AFM
chains have shorter coherence lengths than FM ones. Parameters:
(\(\Delta\), \(\alpha\)) = (0.3, 0.2)t. \label{fig:coh_len_afm_fm}}
\end{figure}
%%%%%%%%%%%%%%%%%%%%%%%%%%%%%%%%%%%%%%%%%%%%%%%%%%%%%%%%%%%%%%%%%%%%%%%%%%%

First we compute the
coherence length \(\xi = v_F/\Delta_{\rm gap}\) in 1D for the AFM and FM
variants of minimal and extended models for $\Delta < t$ (Fig.\,\ref{fig:coh_len_afm_fm}).
Where there are multiple bands crossing the Fermi surface, yielding multiple
Fermi velocities \(v_F\), we choose the largest \(v_F\). Here \(\Delta_{\rm gap}\)
indicates the topological gap, \emph{i.e.}, the effective gap size in the topological phase.
Note that this definition is necessarily approximate; the exact definition of $\xi$ contains
additional factors which we neglect because we are interested in overall trends
in the coherence length, not the precise value. For the minimal model we find that in the
nontrivial phase the mean coherence length for the AFM variant is about seven times smaller
than for the FM variant (where all other parameters are identical). Similarly, for the extended
model the mean coherence length in the nontrivial phase for the AFM variant is about four
times smaller than for the FM variant. For both models this is due to both a
larger gap size and smaller Fermi velocity. This implies that MZMs in AFM chains should be
much more localized than for FM chains; indeed we find this to be the case.

\begin{figure}
\centering
\includegraphics{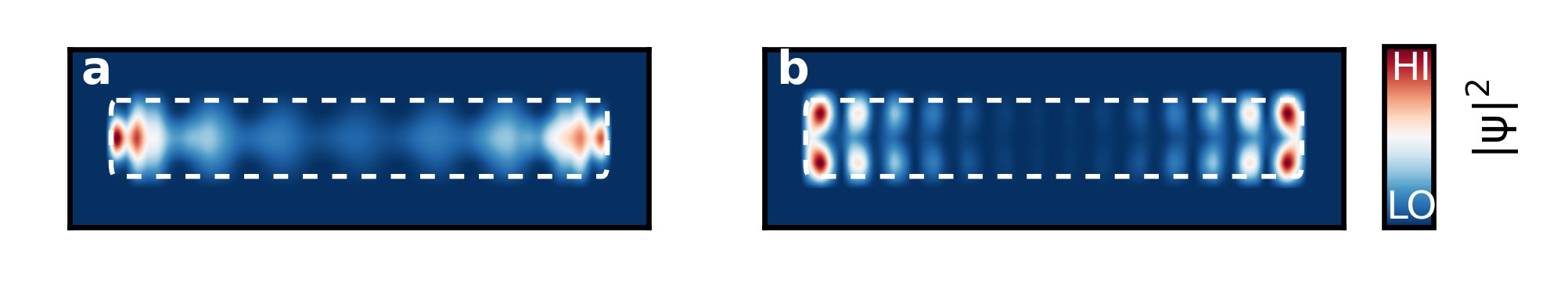}
\caption{Side features in the AFM extended model. (a) MZMs
are found at the ends of the chains and on the adatoms, with some decay into
the bulk. Parameters: (\(\mu\), \(J\)) = (5, 3.5)t. (b) For few fine-tuned situations, MZMs are more local to the sides at chain ends, reminiscent of the side features\,\citep{crawford_majorana_2021} and the double eye feature\,\cite{feldman_high-resolution_2017}. Parameters: (\(\mu\),
\(J\)) = (1, 3.5)t. Parameters for both panels: (\(\Delta\), \(\alpha\),
\(L\)) = (0.3t, 0.2t, 24). \label{fig:side_features}}
\end{figure}

We now look for side features. First we consider the minimal model, with either FM chains or
AFM chains on an extended 2D substrate. For FM chains hybridized MZMs are easily found due to
the long coherence lengths, but their spectral weight is usually distributed along the chain.
In some cases the spectral weight of the hybridized MZMs is found distributed across both chain
and substrate. For AFM chains we can find hybridized MZMs for only a few fine-tuned parameters;
in these cases the spectral weight is similar to that of FM chains.
The absence of side features in the minimal model is not unexpected, because the single-orbital site is coupled to both magnetic Zeeman field and superconducting pairing term, and the hopping structure between these single-orbital sites is too simple.
For the AFM variant of the extended model the spectral weight is always
found at chain ends, and only for a small range of parameters do we find the
spectral weight on the substrate beside the chain (see Fig.\,\ref{fig:side_features}).
This is reminiscent of features found in the FM
variant\,\citep{crawford_majorana_2021} and of the double-eye feature in
Ref.\,\citep{feldman_high-resolution_2017}. We emphasize that the
Fe/Pb(110) system described in the latter reference of course involves a
different unit cell with FM couplings and so does not directly
correspond to our results presented here.

We conclude that in general the spectral gap is larger and the Fermi velocity smaller
for AFM chains compared to FM, all other factors equal. This means that
coherence lengths are shorter and hybridization is suppressed; we
do not find side features in our AFM models. Thus AFM chains are a candidate
for more localized and stable MZMs.

%%%%%%%%%%%%%%%%%%%%%%%%%%%%%%%%%%%%%%%%%%%%%%%%%%%%%%%%%%%%%%%%%%%%%%%%%%%
%%%%%%%%%%%%%%%%%%%%%%%%%%%%%%%%%%%%%%%%%%%%%%%%%%%%%%%%%%%%%%%%%%%%%%%%%%%

%%%%%%%%%%%%%%%%%%%%%%%%%%%%%%%%%%%%%%%%%%%%%%%%%%%%%%%%%%%%%%%%%%%%%%%%%%%
%
%
%%%%%%%%%%%%%%%%%%%%%%%%%%%%%%%%%%%%%%%%%%%%%%%%%%%%%%%%%%%%%%%%%%%%%%%%%%%
\hypertarget{n-leg-ladders}{%
\section{\texorpdfstring{\(N\)-leg
ladders}{N-leg ladders}}\label{n-leg-ladders}}

\begin{figure}[t]
\centering
\includegraphics{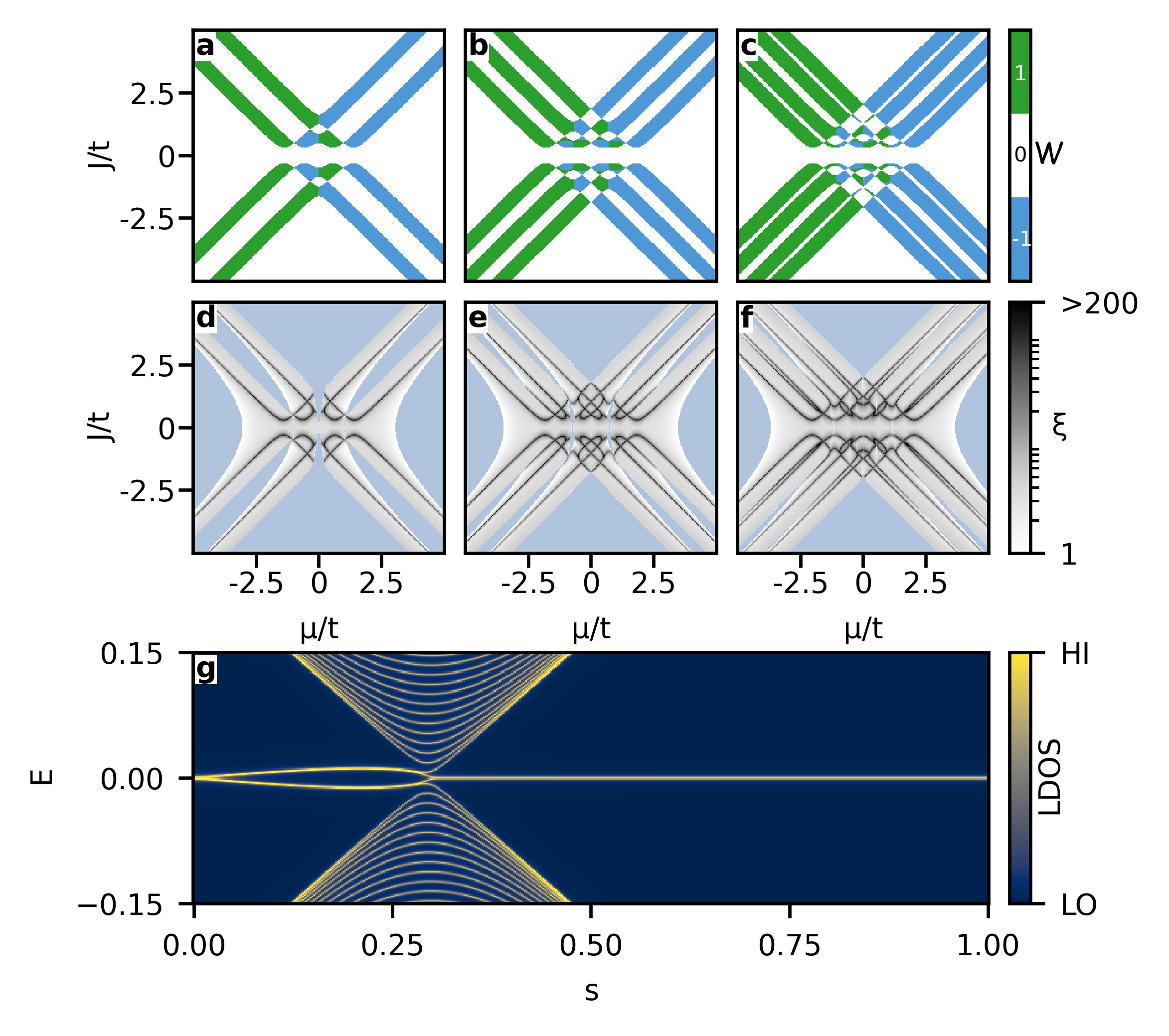}
	\caption{Topological phase diagrams (a-c) and coherence length diagrams (d-f) for (a,d) two-leg, (b,e) three-leg, and (c,f) four-leg ladders, in the minimal model. Blue regions have undefined coherence length, \emph{i.e.} a gapped normal state. (g) Spectra for a two leg ladder with inter-chain couplings scaled by \(s\). At \(s\approx0.3\) there is a phase transition; to the left of this there are four MZMs ($\mathcal{W}=2$) and to the right there are two MZMs ($\mathcal{W}=1$).  (a-f) Parameters: (\(\Delta, \alpha\)) = \((0.3, 0.2)t\); (g) parameters: \((\Delta, \alpha, \mu, J, L)\) = \((1.2t, 0.8t, 3.5t, 2.5t, 128)\).
\label{fig:n-leg_ladders}}
\end{figure}

A natural experimental extension from single chains (\emph{i.e.}, a single row
of adatoms) is to deposit two or more rows of adatoms on the
superconducting surface, each row alongside each other. In reference to
extensions of spin chains we call these \emph{\(N\)-leg ladders}, with $N$ the number of rows.
We study two-, three-, and four-leg ladders using the minimal model and without a substrate;
this is sufficient to elucidate the essential physics.
We find that the topological phase is larger for $N$-leg ladders than for single AFM chains (Fig.\,\ref{fig:n-leg_ladders}a-c),
and the size of the topological phase increases with the number of legs; 
we stress that this trend only holds for a few rows of adatoms, before
the system becomes too close to the 2D limit. In 2D with adatoms covering
the entire surface the system is known to be gapless due to nodal points
in the spectrum (see Appendix\,\ref{minimal-model-spectra-and-gap-closing}3 for details).
We find that MZMs in these topological phases look essentially identical
to those found in single chains, including an exponential decay into the ladders.
Thus hybridization is a potential issue and so we compute coherence lengths (Fig.\,\ref{fig:n-leg_ladders}d-f).
We find
(a) the spectral gap for these ladders is comparable to single chains, and
(b) the mean coherence length in all cases is similar to that of single chains in the previous section.
Note that the phase boundaries are visible in these coherence length plots 
because at the gap closing points \(\Delta_{\rm gap}\) becomes vanishingly small,
leading to a diverging coherence length \(\xi \sim \Delta_{\rm gap}^{-1}\).

A closer inspection of the phase diagrams in Fig\,\ref{fig:n-leg_ladders}a-c reveals
that after fixing parameters adding a second chain to a single-chain system (and similarly, adding additional chains to an $N$-leg ladder)
may change the topological phase. This might open up the exciting opportunity 
for experimentalists to convert a topologically trivial single chain into a
topologically nontrivial two-leg ladder. In particular, this is feasible 
for experimental systems built with single-atom manipulation techniques. 
We investigate how the transition from a single chain to a two-leg ladder 
occurs by modifying the inter-chain couplings (\emph{i.e.} the hoppings and Rashba SOC)
by a global scale factor $s \in [0,1]$. In Fig.\,\ref{fig:n-leg_ladders}\,g we show
the spectrum as a function of $s$, where $s=0$ corresponds to the two decoupled 
chains and $s=1$ corresponds to a normal two-leg ladder.
At $s=0$ we find a $\mathcal{W}=2$ phase with four MZMs localized to the ends of the two decoupled chains,
while at $s=1$ we find a $\mathcal{W}=1$ phase with only two MZMs, localized at the ends of the ladder. 
In between there is a topological phase transition (revealed by a gap closing) at $s^\ast\approx 0.3$. 
In fact, for $N$ decoupled chains we find for the total system $\mathcal{W}=N$ (\emph{i.e.}, $N$ pairs of MZMs). These MZMs hybridize when $s$ is small (and thus the coupling weak). At some finite value $s^\ast$ the topological phase transition occurs and for $s > s^\ast$ one finds either a topological phase with $\mathcal{W}=1$, and one pair of MZMs, or a $\mathcal{W}= 0 \mod 2$ phase with an even number of MZM pairs, depending on the system parameters.

%%%%%%%%%%%%%%%%%%%%%%%%%%%%%%%%%%%%%%%%%%%%%%%%%%%%%%%%%%%%%%%%%%%%%%%%%%%
%%%%%%%%%%%%%%%%%%%%%%%%%%%%%%%%%%%%%%%%%%%%%%%%%%%%%%%%%%%%%%%%%%%%%%%%%%%

Naturally one can imagine a zoo of $N$-leg ladders, depending on if the surface is purely FM,
row-wise AFM, Néel AFM, and so on. In these cases the nature of the topological phases 
depends on the details. For example, a purely-FM two-leg ladder behaves essentially the
same as discussed previously, while a two-leg ladder on a Néel surface has only $|\mathcal{W}|=0,2$ 
phases and so has no MZMs. On the other hand, a \emph{three}-leg ladder on a Néel surface
has $|\mathcal{W}|=1$ phases, and we see a transition from six MZMs to two MZMs,
similar to that shown in Fig.\,\ref{fig:n-leg_ladders}\,g.

We conclude that \(N\)-leg ladders offer a way to convert topologically trivial
single AFM chains into nontrivial systems, simply be adding an extra row of adatoms.
This is experimentally tractable, and overcomes the small topological phases 
in single AFM chains due to the dependence on Rashba SOC magnitude.
MZMs in \(N\)-leg ladders have similar coherence lengths to single AFM chains,
and so there is no cost to the stability of MZMs using this technique.

%%%%%%%%%%%%%%%%%%%%%%%%%%%%%%%%%%%%%%%%%%%%%%%%%%%%%%%%%%%%%%%%%%%%%%%%%%%
%
%
%%%%%%%%%%%%%%%%%%%%%%%%%%%%%%%%%%%%%%%%%%%%%%%%%%%%%%%%%%%%%%%%%%%%%%%%%%%
\hypertarget{t-junctions-on-an-extended-substrate}{%
\section{T-junctions on an extended
substrate}\label{t-junctions-on-an-extended-substrate}}

A single MSH chain, which can host at most one MZM per chain end, is insufficient to implement a topological qubit ---
which requires either 3 or 4 MZMs
\citep{nayak_non-abelian_2008, sarma_majorana_2015,alicea_non-abelian_2011} --- or to perform braiding. When MZMs are constrained to move in one dimension, such as on
a chain, then obviously they cannot be exchanged without passing through
(and hybridizing or fusing with) each other. Hence a realistic
topological quantum computer based on MSH chains must involve networks
of coupled chains. The fundamental unit of such networks is the
\emph{tri-junction}\,\citep{alicea_non-abelian_2011}. A tri-junction is
comprised of three chains connected at a single point. The system is
arranged so that there are MZMs at the ends of two chains, and the third
is an auxiliary where one of the MZMs can be ``parked'' while the other MZM changes its position. On a square lattice
two of the chains are in the same direction and so the geometry is
called a \emph{T-junction}.
We note that on a hexagonal lattice, one would consider a \emph{Y-junction} instead.

\begin{figure}
\centering
\includegraphics{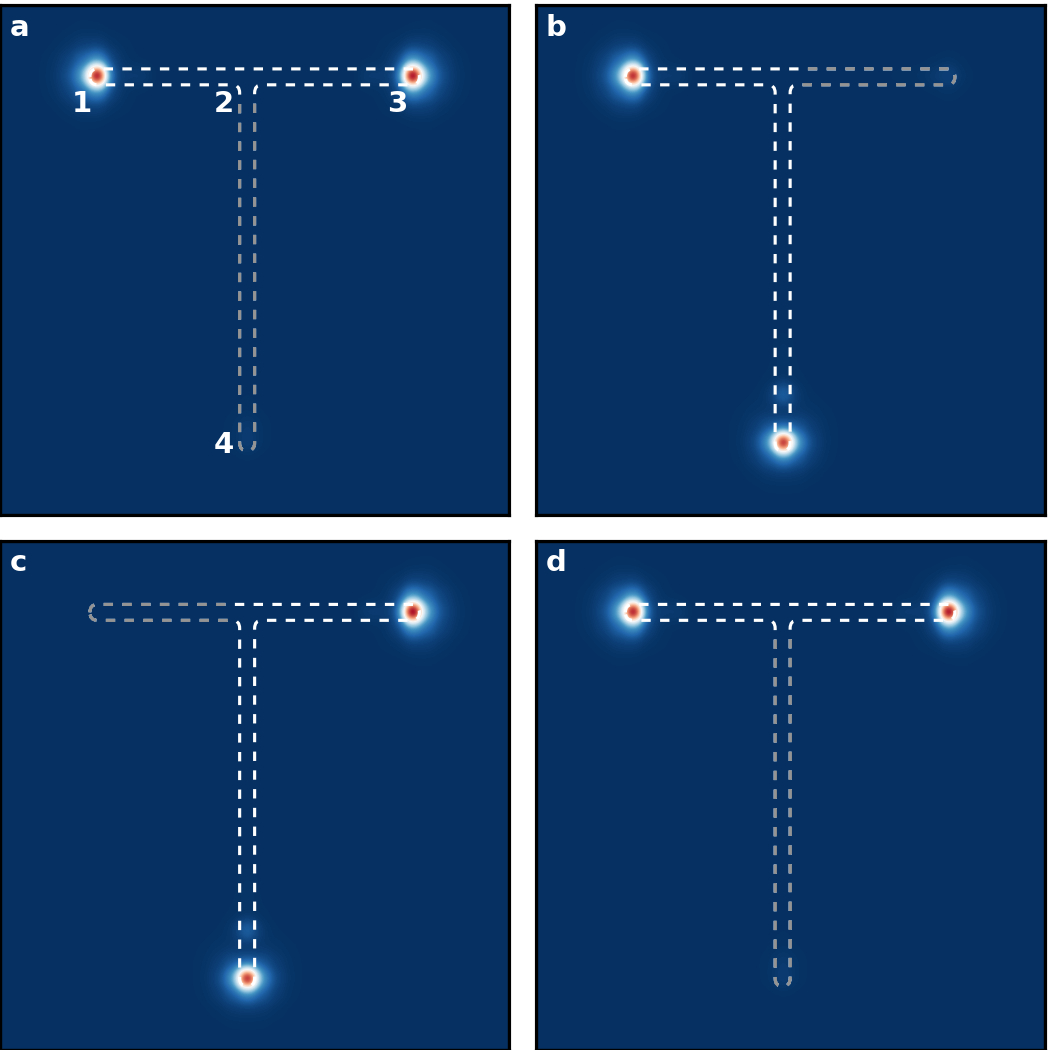}
\caption{Braiding of two MZMs in a T-junction (for the minimal model). Each
panel shows the zero-energy LDOS. MZMs are moved by varying a local chemical potential, tuning the system from topological to trivial. Dashed lines indicate magnetic adatoms, with gray regions corresponding to the trivial phase.
By applying the sequence
of local potentials shown, MZMs at initial positions 1 and 3 can be
exchanged. Parameters: (\(\Delta\), \(\alpha\),
\(\mu_{\rm global}\), \(\mu_{\rm local}\), \(J\), $L_x$, $L_y$) = (1.2t, 0.8t, 3.75t, 5t,
3.5t, 11, 24). \label{fig:t-junction-braid}}
\end{figure}

Because the tri-junction was introduced in the context of Rashba nanowires with an applied \emph{uniform}
external field, it is important to verify that the principle similarly works with
MSH networks involving a mixture of FM and AFM couplings.
We show an example of this in
Fig.\,\ref{fig:t-junction-braid}, computed statically (\emph{i.e.}, no
quantum dynamics is involved). We include in the supplement a short
animation showing snapshots of the full braiding process. Depending on
parameters, MZMs may be found at any of the locations 1-4 as marked in
Fig.\,\ref{fig:t-junction-braid}\,a. For clarity we study the minimal model
with a large gap; the calculations can be repeated for a small gap but
this requires long chains to overcome hybridization. We apply a global
chemical potential \(\mu_{\rm global}\) everywhere, and also a local
potential \(\mu_{\rm local}\) on chain sites.
As described in Ref.\,\citep{alicea_non-abelian_2011}, by applying a
``keyboard'' of gates along a chain a tunable domain wall can be
engineered to move MZMs and hence braid them. The chain between 1 and 3
is antiferromagnetic, and the chain between 2 and 4 is ferromagnetic: we
emphasize not only the significance of the mixed magnetic orders, but
also that these calculations include the full substrate.
In the particular example [Fig.\,\ref{fig:t-junction-braid}\,a], we start with MZMs at 1 and 3. We change
the local chemical potential and move the MZM from 3 to 4. Applying a
local chemical potential between 1 and 2 moves the MZM from 1 to 3, and
then reapplying the initial local chemical potential moves the MZM from
4 (initially at 3) to 1. This completes the braid. In principle we can
include other effects such as disorder but leave that for future
work. Note that we choose large gap parameters for Fig.\,\ref{fig:t-junction-braid}, which minimizes
hybridization, for clarity. The same procedure works for small gaps but
requires much longer chains.

%%%%%%%%%%%%%%%%%%%%%%%%%%%%%%%%%%%%%%%%%%%%%%%%%%%%%%%%%%%%%%%%%%%%%%%%%%%
%%%%%%%%%%%%%%%%%%%%%%%%%%%%%%%%%%%%%%%%%%%%%%%%%%%%%%%%%%%%%%%%%%%%%%%%%%%

\begin{figure}
\centering
\includegraphics{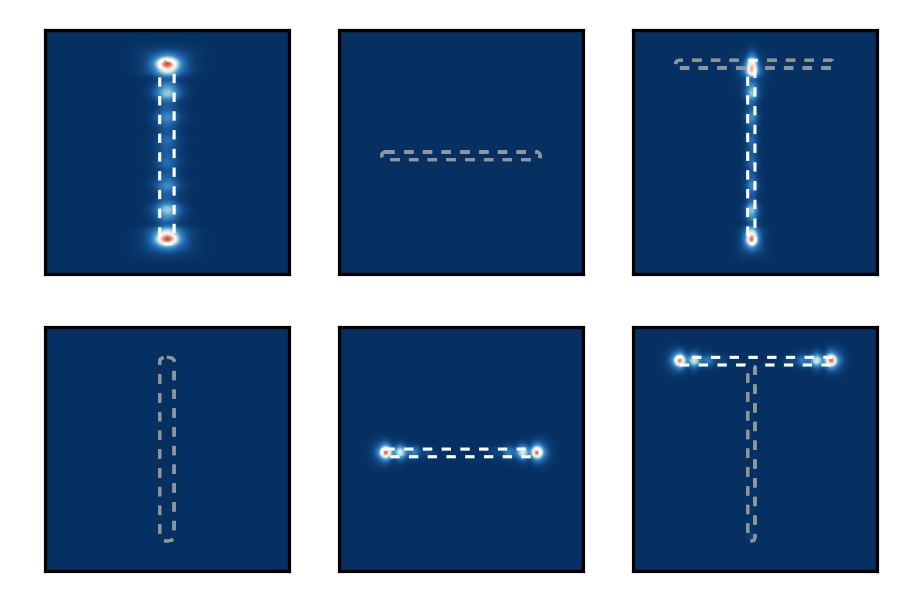}
\caption{State initialization by tuning the global chemical potential. All
panels show the zero-energy LDOS for the same $\mu\equiv \mu_{\rm global}$ as indicated on
the left. (a) \(\mu\) is chosen such that the FM part of the T-junction is
in the nontrivial phase and the AFM part is in the trivial phase. Hence MZMs
are found at positions 2 and 4. (b) \(\mu\) is chosen such that the
phases are the opposite way around and the MZMs are found at
positions 1 and 3. Parameters: (\(\Delta\), \(\alpha\), \(J\)) = (0.3,
0.2, 3.25)t. \label{fig:t-junction}}
\end{figure}

We can exploit the fact that an AFM chain is only in a topological phase
for small range of parameters compared to the FM chain to select which
position in a T-junction MZMs may be found. Fig.\,\ref{fig:t-junction}
shows two different MZM configurations, corresponding to two different
global chemical potentials. In (a) MZMs can be initialized at positions
2 and 4 by combining an FM chain in the nontrivial phase with an AFM
chain in the trivial. In (b) the MZMs are found at positions 1 and 3 by
flipping the FM chain into a trivial phase and the AFM chain into a
nontrivial phase. This may be useful for state initialization: Consider
a network of chains consisting of discrete T-junction units, with some
coupling between these units. By gating each T-junction individually, MZMs can be
initialized in pairs at well-defined positions. Local gates can then be used to
move the MZMs, as described above.

%%%%%%%%%%%%%%%%%%%%%%%%%%%%%%%%%%%%%%%%%%%%%%%%%%%%%%%%%%%%%%%%%%%%%%%%%%%
%%%%%%%%%%%%%%%%%%%%%%%%%%%%%%%%%%%%%%%%%%%%%%%%%%%%%%%%%%%%%%%%%%%%%%%%%%%

%%%%%%%%%%%%%%%%%%%%%%%%%%%%%%%%%%%%%%%%%%%%%%%%%%%%%%%%%%%%%%%%%%%%%%%%%%%
%
%
%%%%%%%%%%%%%%%%%%%%%%%%%%%%%%%%%%%%%%%%%%%%%%%%%%%%%%%%%%%%%%%%%%%%%%%%%%%
\hypertarget{discussion}{%
\section{Discussion}\label{discussion}}

Magnetic chains on the surface of a superconductor may have FM or AFM
coupling (or more exotic magnetic ordering tendencies) depending on the crystal direction. We know from simplified models
that these chains may be topological superconductors with MZMs localized
at the ends. Experiments are often ambiguous so it is essential to
develop robust theory for more realistic systems, \emph{i.e.}, models
involving several superconducting atoms and orbitals, separate from the
magnetic adatom, with small ($\Delta \ll t$ or at least $\Delta < t$) superconducting gap.
Ideally this involves constructing a tight-binding model from \emph{ab
initio} methods on a slab, although this remains computationally
expensive.

Unlike FM chains, it turns out for our models of AFM chains that the topological phase is
closely related to the magnitude of Rashba spin-orbit coupling; this
occurs for both the minimal and the extended models. By the usual energy
scales \(\alpha < \Delta < t\) and so in realistic systems the
topological phase may only be found for a very small range of
parameters. We conjecture that this is true in general for all AFM
chains.

MZMs at AFM chain ends may have some advantages over the FM case. For both
models we find that in general AFM chains have a (much) larger topological
gap than the FM chains, while keeping all other parameters equal. This means that
coherence lengths are shorter for AFM chains and so hybridization is
suppressed. Although we find that the minimal model is too simple to
show side features, we can compare the AFM variant of the extended model
to the FM variant and find that there are no side features
(Fig.\,\ref{fig:side_features}) due to the suppressed hybridization, except for very few fine-tuned parameter points.

We propose that topological phases may be found over a greater range of
parameters by depositing two or more rows of adatoms alongside each
other, forming two-and-three-leg ladders. As long as these systems are
sufficiently one-dimensional we find topological phases and MZMs as in
the previous cases. There appears to be no advantage in localization:
these \(N\)-leg ladders have much the same gap sizes as chains.
Constructing an \(N\)-leg ladder on a superconducting  substrate using single-atom manipulation should be
experimentally feasible.

Because a topological qubit may be constructed from a network of chains
it is important to test if anything unexpected occurs when AFM chains
couple with FM chains. We examine a fundamental unit of 1D TSC networks,
the T-junction.
If each chain is gated in a keyboard-like fashion
then MZMs can be moved by engineering a domain wall in different parts
of the junction. We find that the mixed magnetic ordering has no
obvious disadvantage and a combined AFM/FM T-junction can be used to braid
two MZMs. Of course hybridization can be an issue for MSH chains and so
any experimental realization of this unit must be sufficiently large and
sufficiently adiabatic to yield good results. We also propose that
qubits can be initialized by tuning the global chemical potential.
Because AFM chains are only in a nontrivial phase for a smaller range of
parameters than FM chains, then by tuning the global chemical potential
we can choose to have the AFM part of a T-junction in the trivial phase
and the FM part in the nontrivial phase, or vice versa. In this fashion
we can initialize MZMs in different parts of a T-junction.

%%%%%%%%%%%%%%%%%%%%%%%%%%%%%%%%%%%%%%%%%%%%%%%%%%%%%%%%%%%%%%%%%%%%%%%%%%%
%
%
%%%%%%%%%%%%%%%%%%%%%%%%%%%%%%%%%%%%%%%%%%%%%%%%%%%%%%%%%%%%%%%%%%%%%%%%%%%
\hypertarget{conclusion}{%
\section{Conclusion}\label{conclusion}}

MSH chains are a promising platform for engineering 1D topological
superconductors; much experimental and theoretical progress has been
made towards realizing MZMs on this platform. Most work has focused on
chains with ferromagnetic order, while other magnetic orders are largely
unexplored. The recent demonstration of antiferromagnetic Mn chains on
Nb(110) motivates us to investigate the topological properties of AFM
chains. We have studied two models inspired by the Nb(110) surface: a
minimal model on a square lattice with AFM order in one direction and FM
in the other; and an extended model with a unit cell containing separate
superconducting and magnetic sites, with AFM coupling between magnetic
sites. We focus on small gaps and explicitly include the superconducting
substrate. We find topological phases for both models, with robust MZM
end-states. We find for both models the topological gap is larger for
AFM models than for FM models and MZMs are less likely to hybridize when
chains are short, compatible with recent experimental results.
We find that
the topological phase is closely related to the size of the Rashba
amplitude and for realistic systems the topological phase is only a
small part of parameter space. This could be improved by engineering
\(N\)-leg ladders, comprised of several adjacent chains of adatoms. We
finally consider T-junctions comprised of an AFM and
a FM chain, as naturally occurs on the Nb(110) surface for Mn
adatoms. We show that braiding %via local chemical potentials
is possible, and states can be initialized via tuning the global
chemical potential to flip the antiferromagnetic chain in and out of the
topological phase.

\begin{acknowledgements}
S.R.\ acknowledges support from the Australian Research Council through Grant No.\ DP200101118. D.M.\ acknowledges support by the U.\ S.\ Department of Energy, Office of Science, Basic Energy Sciences, under Award No.\ DE-FG02-05ER46225. R.W.\ gratefully acknowledges funding by the Cluster of Excellence ‘Advanced Imaging of Matter’ (EXC 2056 - project ID 390715994) of the Deutsche Forschungsgemeinschaft (DFG), by the DFG via SFB 925 – project ID 170620586 and by the European Union via the ERC Advanced Grant ADMIRE (project No.\ 786020).
\end{acknowledgements}

\appendix

\hypertarget{minimal-model-spectra-and-gap-closing}{%
\section{Minimal model spectra and gap
closing}\label{minimal-model-spectra-and-gap-closing}}

In the main text we analyze the bulk topological properties of an AFM
(FM) chain by taking the 1D variant of the minimal model, \emph{i.e.}, by
dropping \(y\) (\(x\)) terms and setting \(\Lambda = \Lambda^*\). We
present here the spectra and gap closing conditions for these cases. We
also consider the minimal model in 2D under the dense limit
\(\Lambda^* = \Lambda\), and show spectra and gap closing conditions.
This coincides with the case of an \(N\)-leg ladder where
\(N\sim L\), \(L\) being the chain length.

For all variants \(\tau_{x,y,z}\) are the Pauli matrices acting on
particle-hole space; \(\nu_{x,y,z}\) are Pauli matrices acting on
sublattice space; and \(\sigma_{x,y,z}\) are the Pauli matrices acting
on spin space. \(\sigma_0, \tau_0\) and $\nu_0$ are the identity matrices. Tensor products are implicit between each Pauli
matrix and multiply right-to-left.

\hypertarget{d-afm-variant}{%
\subsection{1D AFM variant}\label{d-afm-variant}}

By imposing PBC the momentum space Hamiltonian is

\begin{equation}
\nonumber
H =  \frac{1}{2} \sum_k \psi_k^\dagger \mathcal{H}_k \psi_k
\end{equation}
with the Bloch matrix
\begin{align}
\mathcal{H}_k = &\ \varepsilon_k \tau_z\nu_x\sigma_0 + \mu \tau_z\nu_0\sigma_0 \\
&+ \alpha_k \tau_0\nu_x\sigma_x + J \tau_z \nu_z \sigma_z + \Delta \tau_y\nu_0\sigma_y ,
\end{align}
where
\begin{align*}
\varepsilon_k &= 2t \cos (k/2)\ , \\
\alpha_k &= 2 \alpha \sin (k/2)\ . \\
\end{align*}
We use the basis
\begin{equation*}
\psi_k = (a_{k,\uparrow}, a_{k,\downarrow},b_{k,\uparrow},b_{k,\downarrow},a_{-k,\uparrow}^\dagger,a_{-k,\downarrow}^\dagger,b_{-k,\uparrow}^\dagger,b_{-k,\downarrow}^\dagger) .
\end{equation*}
The energy dispersion is given by
\begin{align}
\nonumber
E_k = \pm \Big[&(\alpha_k + \delta_\pm \mu)^2 + \varepsilon_k^2 + \Delta^2 +J^2  \\
&\pm 2 \sqrt{(\alpha_k + \delta_\pm \mu)^2 \varepsilon_k^2 + J^2 ((\alpha_k + \delta_\pm \mu)^2 + \Delta^2)}\Big]^{1/2} .
\end{align}

Here \(\delta_\pm = \pm 1\). This gives the gap closing condition
\(J^2 = (2\alpha \pm \mu)^2 + \Delta^2\). Note here how
the overall scale of the topological phase depends on \(\alpha\). Hence
when \(\alpha\) is small due to the energy scale \(\alpha < \Delta\),
the topological phase will be small.

\hypertarget{d-fm-variant}{%
\subsection{1D FM variant}\label{d-fm-variant}}

By imposing PBC the momentum space Hamiltonian is

\begin{equation}
    H = \frac{1}{2} \sum_k \psi_k^\dagger \mathcal{H}_k \psi_k
\end{equation}
with the Bloch matrix
\begin{align}
\mathcal{H}_k = &\ \varepsilon_k \tau_z\sigma_0 + \alpha_k \tau_0\sigma_y + J \tau_z  \sigma_z + \Delta \tau_y\sigma_y ,
\end{align}
where
\begin{align*}
\varepsilon_k &= 2t \cos (k) - \mu\ , \\
\alpha_k &= 2 \alpha \sin (k)\ . \\
\end{align*}
We use the basis
\begin{equation*}
\psi_k = (a_{k,\uparrow}, a_{k,\downarrow},a_{-k,\uparrow}^\dagger,a_{-k,\downarrow}^\dagger) .
\end{equation*}
The energy dispersion is given by
\begin{align}
\nonumber
E_k = \pm \Big[&J^2 + \Delta^2 + \varepsilon_k^2 + \alpha_k^2  \\
&\pm 2 \sqrt{J^2(\Delta^2 + \varepsilon_k^2) + \varepsilon_k^2 \alpha_k^2}\Big]^{1/2} .
\end{align}

This gives the gap closing condition
\(J^2 = (2t)^2 + \Delta^2\). Hence for FM chains only the presence of Rashba
spin-orbit coupling is required to give a topological phase; the
magnitude is not important.

\hypertarget{dense-variant}{%
\subsection{Dense variant}\label{dense-variant}}

Here we study the minimal model in 2D with magnetic atoms at every site \emph{i.e.}, the dense impurity limit. This corresponds to a row-wise antiferromagnetic structure plus superconductivity. By imposing PBC the momentum space Hamiltonian is
\begin{equation}
    H =  \frac{1}{2} \sum_k \psi_k^\dagger \mathcal{H}_k \psi_k
\end{equation}
with the Bloch matrix
\begin{align}
\nonumber
\mathcal{H}_k = &\ \varepsilon_{k_x} \tau_z\nu_x\sigma_0 + \varepsilon_{k_y} \tau_z\nu_0\sigma_0
\label{eq:dense_ham}
+ \alpha_{k_x} \tau_0\nu_x\sigma_x + \alpha_{k_y} \tau_0\nu_0\sigma_y \\
&+ J \tau_z \nu_z \sigma_z + \Delta \tau_y\nu_0\sigma_y ,
\end{align}
where
\begin{align*}
\varepsilon_{k_x} &= 2t \cos (k_x/2)\ , \\
\alpha_{k_x} &= 2 \alpha \sin (k_x/2)\ , \\
\varepsilon_{k_y} &= 2t \cos (k_y) - \mu\ , \\
\alpha_{k_y} &= 2 \alpha \sin (k_y)\ . \\
\end{align*}
We use the basis
\begin{equation*}
\psi_k = (a_{k,\uparrow}, a_{k,\downarrow},b_{k,\uparrow},b_{k,\downarrow},a_{-k,\uparrow}^\dagger,a_{-k,\downarrow}^\dagger,b_{-k,\uparrow}^\dagger,b_{-k,\downarrow}^\dagger) .
\end{equation*}
The energy dispersion is too complicated to write down here, but there
are several useful limits:
\begin{align*}
\begin{array}{c|c|c|c|c}
& \vec{k} = (0,0) & \vec{k} = (0,\pi) & \vec{k} = (\pi,0) & \vec{k} = (\pi,\pi)  \\ \hline
\varepsilon_{k_x} & 2t & 2t & 0 & 0\\
\varepsilon_{k_y} & 2t-\mu & -2t-\mu & 2t-\mu & -2t-\mu  \\
\alpha_{k_x} & 0 & 0 & 2\alpha & 2\alpha\\
\alpha_{k_y} & 0 & 0 & 0 & 0  \\ \hline
\end{array}
\end{align*}

These imply the spectra \(E_k^{k_x=0}\) and \(E_k^{k_x=\pi}\),

\begin{align}
\nonumber
E_k^{k_x=0} &= \pm \Big[\varepsilon_{k_x}^2 + \varepsilon_{k_y}^2 + J^2 + \Delta^2 \\
&~~~~~~~\pm \sqrt{\varepsilon_{k_y}^2(\varepsilon_{k_x}^2+J^2) + J^2\Delta^2} \Big]^{1/2},  \\[10pt]
E_k^{k_x=\pi} &= \pm \alpha_{k_x} \pm J \pm \sqrt{\varepsilon_{k_y}^2 + \Delta^2} .
\end{align}

There is no gap closing associated with \(E_k^{k_x=0}\).
\(E_k^{k_x=\pi}\) implies the gap closing condition
\(J^2 = (\pm \alpha_{k_x} \pm \sqrt{\varepsilon_{k_y}^2+\Delta^2})^2\),
evaluated at \(\vec{k}=(\pi,0),(\pi,\pi)\). Hence \(N\)-leg ladders are
only valid for small enough \(N\) to avoid these \(\vec{k}\) points,
which close the gap and hence destroy the topological phase.

\hypertarget{extended-model-definition}{%
\section{Extended model definition}\label{extended-model-definition}}

Here we define the hopping parameters used in the extended model. Table\,\ref{tab:extended_model} lists the elements of the tensor \(t_{ij}^{a b}\)  used in Eq\,\ref{eq:ham2}, with $t$ the hopping amplitude between atoms $a$ and $b$. \(|i-j| = 0\) indicates that the hopping is onsite and \(|i-j| = 1\) indicates that the hopping is to first neighbor.
\(\alpha_{ij}^{ab}\) is identical, except that each entry in the \(t\) row is replaced with the variable \(\alpha\).

%%%%%%%%%%%%%%%%%%%%%%%%%%%%%%%%%%%%%%%%%%%%%%%%%%%%%%%%%%%%%%%%%%%%%%%%%%%
\begin{table}[t]
\caption{Hopping parameters for the extended model. \(t\) indicates the hopping amplitude. \(a\) and \(b\) indicate which atoms the hopping is between. \(|i-j| = 0\) indicates the hopping is within the unit cell and \(|i-j| = 1\) indicates the hopping is to first neighbor. \(\alpha_{ij}^{ab}\) is identical to this table, except that the \(t\) row is constant.
}\label{tab:extended_model}
\begin{tabular*}{\columnwidth}{@{\extracolsep{\fill}}c|cccccccccccccccccccc}
	\(t\) & 3 & 3 & 3 & 3 & 2 & 2 & 1 & 1 & 1 & 1 & 1 & 1 & 1 & 1 & 1 & 1 & 0.2 & 0.2 & 0.2 & 0.2 \\ \hline
	\(a\) & 1 & 0 & 5 & 4 & 2 & 6 & 1 & 5 & 1 & 1 & 2 & 3 & 5 & 5 & 6 & 7 & 2 & 3 & 6 & 7\\
	\(b\) & 0 & 5 & 4 & 1 & 3 & 7 & 5 & 1 & 2 & 3 & 5 & 5 & 6 & 7 & 1 & 1 & 6 & 7 & 2 & 3\\
	\(|i-j|\) & 0 & 0 & 0 & 1 & 0 & 0 & 0 & 1 & 0 & 0 & 0 & 0 & 0 & 0 & 1 & 1 & 0 & 0 & 1 & 1\\
\hline\hline
\end{tabular*}
\end{table}
%%%%%%%%%%%%%%%%%%%%%%%%%%%%%%%%%%%%%%%%%%%%%%%%%%%%%%%%%%%%%%%%%%%%%%%%%%%

\newpage

\bibliography{paper.bib}

\end{document}